\documentclass[10pt]{IEEEtran}

\def\BibTeX{{\rmfamily B\kern-.05em{\scshape i\kern-.025em b}\kern-.08em \TeX}}

\usepackage{cite}
\ifCLASSINFOpdf
\else
\fi

\usepackage[cmex10]{amsmath}

\usepackage[dvips]{graphicx}

\begin{document}

\title{Traffic and Statistical Multiplexing
Characterization of 3D Video Representation
 Formats\\ (Extended Version)\thanks{Technical Report, School of 
Electrical, Computer, and Energy Eng., Arizona State Univ., November 2012.
This extended technical report accompanies~\cite{PuSRK13}.}
\thanks{Supported in part by the 
National Science Foundation
  through grant No.\ CRI-0750927.}
  \thanks{Please direct correspondence to M.~Reisslein.}}

\author{Akshay~Pulipaka, Patrick~Seeling, Martin~Reisslein, and
Lina J.~Karam
\thanks{A.~Pulipaka, M.~Reisslein, and L.J.~Karam
are with the School of Electrical, Computer, and Energy Engineering
Arizona State University, Tempe, AZ 85287-5706,
\texttt{http://trace.eas.asu.edu},
Email: \texttt{\{akshay.pulipaka, reisslein, karam\}@asu.edu}}
\thanks{P.~Seeling is with Central Michigan University,
Mount Pleasant, MI 48859, Email: \texttt{pseeling@ieee.org}}}

\maketitle

\begin{abstract}
The network transport of 3D video, which contains
two views of a video scene, poses significant challenges due to the increased
video data compared to conventional single-view video.
Addressing these challenges requires a thorough understanding of the
traffic and multiplexing characteristics
of the different representation formats of 3D video.
We examine the average bitrate-distortion (RD) and
bitrate variability-distortion (VD) characteristics of three main
representation formats.
Specifically, we compare
multiview video (MV) representation and encoding,
frame sequential (FS) representation, and side-by-side (SBS) representation,
whereby conventional single-view encoding is employed for the
FS and SBS representations.
Our results for long 3D videos in full HD format indicate that the
MV representation and encoding achieves the highest RD efficiency,
while exhibiting the highest bitrate variabilities.
We examine the impact of these bitrate variabilities on
network transport through extensive statistical multiplexing simulations.
We find that when multiplexing a small number of streams,
the MV and FS representations require the same bandwidth.
However, when multiplexing a large number of streams or smoothing traffic,
the MV representation and encoding reduces the
bandwidth requirement relative to the FS representation.
\end{abstract}

\IEEEpeerreviewmaketitle

\section{Introduction}
Multiview video provides several views taken from
different perspectives, whereby each view consists of a sequence of
video frames (pictures).
Multiview video with two
marginally different views of a given scene can be displayed
to give viewers the perception of depth and is therefore
commonly referred to as three-dimensional (3D) video or
stereoscopic video~\cite{MeMW10,MoUA11,VeTMC,VeMPX04,WiUSHL07};
for brevity we use the term ``3D video'' throughout.
Providing 3D video services over transport networks requires
efficient video compression (coding) techniques and transport mechanisms
to accommodate the large volume of video data from the two views
on bandwidth limited transmission links.
While efficient coding techniques for multiview video have
been researched extensively in recent years~\cite{ChWUHLG09,VeWS}, the network
transport of encoded 3D video is largely an open research problem.

Previous studies on 3D video transport have primarily focused on the
network and transport layer protocols
and file formats~\cite{AkTFC07,GuGST,mvv,ScNa11}.
For instance,~\cite{GuGST,ZiMANAK09} examine the extension of
common transport protocols, such as
the datagram congestion control protocol (DCCP), the stream control
transmission protocol (SCTP), and the user datagram protocol (UDP)
to 3D streaming, while the use of two separate Internet Protocol (IP)
channels for the delivery of multiview video is studied in~\cite{ZhHJYYG09}.
Another existing line of research has studied
prioritization and selective transport mechanisms for multiview
video~\cite{KuCT07,WuSY11}.

In this study, we examine the fundamental traffic
and statistical multiplexing characteristics of
the main existing approaches for representing and encoding 3D video
for long (54,000 frames) full HD ($1920 \times 1080$) 3D videos.
More specifically, we consider
$(i)$ multiview video (MV) representation
and encoding, which exploits the redundancies between the two views,
$(ii)$ frame sequential (FS) representation, which merges the two views to
form a single sequence with twice the frame rate and applies conventional
single-view encoding, and
$(iii)$ side-by-side (SBS) representation, which halves the horizontal
resolution of the views and combines them to form a single frame sequence
for single-view encoding.

We find that the MV representation achieves the most efficient encoding, but
generates high traffic variability, which makes statistical
multiplexing more challenging.
Indeed, for small numbers of multiplexed streams, the FS representation
with conventional single-view coding has the same transmission bandwidth
requirements as the MV representation with multiview coding.
Only when smoothing the MV traffic or multiplexing many streams
can transport systems benefit from the more efficient MV encoding.

In order to support further research on the network transport of 3D video,
we make all video traces~\cite{SeRe12} used in this study publicly available
in the video trace library \texttt{http://trace.eas.asu.edu}.
In particular, video traffic
modeling~\cite{AlAA04,AnLS02,FiSB11,LaKo10,ShSZ11} requires
video traces for model development and validation.
Thus, the traffic characteristics of 3D video
covered in this study will
support the nascent research area of 3D video traffic
modeling~\cite{RoCFC10}.
Similarly, video traffic management mechanisms for a wide range of
networks, including wireless and optical networks,
are built on the fundamental  traffic and multiplexing characteristics
of the encoded video traffic~\cite{AlSA12,DiLL08,QiKo11,SzGi09}.
Thus, the broad traffic and statistical multiplexing evaluations in this
study provide a basis for the emerging research area on
3D video traffic management in transport networks~\cite{CoLAT10,MaDS09}.

\section{Multiview Video Representation, Encoding, and Streaming}
\label{ov}
In this section, we provide a brief overview of the main representation
formats for multiview video~\cite{VeTMC} as well as the applicable
encoding and streaming approaches.

\subsection{Multiview video representation formats}
\label{rep}

With the full resolution multiview format, which we refer to
as multiview (MV) format for brevity,
each view $v,\ v = 1, \ldots, V$,
is represented with the full resolution of the underlying
spatial video format.
For instance, the MV format for the full HD resolution of
$1920\times 1080$ pixels consists of a sequence of
$1920\times 1080$ pixel frames for each view $v$.
Each view has the same frame rate as the underlying temporal
video format.
For example, for a video with a frame rate of $f = 24$ frames/s, each
view has a frame rate of $f = 24$ frames/s.

With the frame sequential (FS) representation, the
video frames of the $V$ views (at the full spatial resolution)
are temporally multiplexed to form a single sequence of video frames
with frame rate $Vf$.
For instance, for $V = 2$ views, the video frames
from the left and
right views are interleaved in alternating fashion to form a single stream
with frame rate $2f$.

Frame-compatible representation formats
have been introduced to utilize the existing
infrastructure and equipment for the transmission of
stereoscopic two-view video~\cite{VeTMC}.
The $V = 2$ views are spatially sub-sampled and
multiplexed to form a single sequence of video frames with
the same temporal and spatial resolution as the underlying video
format~\cite{JVC}.
In the side-by-side (SBS) format, the left and right views are
spatially sub-sampled in the horizontal direction and are then
combined side-by-side.
For instance, for the full HD format, the left and right views are
sub-sampled to $960 \times 1080$ pixels. Thus, when they are combined
in the side-by-side format, they still occupy the full HD resolution for every
frame.
However, each frame contains the left and right views at only half
the original horizontal resolution.
In the top-and-bottom format, the left and right
views are sub-sampled in the vertical direction and combined in
top-and-bottom (above-below) fashion.
For other formats, we refer to~\cite{DKBr11,JVC,LeJH,VeTMC}.
We consider the side-by-side (SBS) representation format in our study,
since it is one of the most widely used frame-compatible
formats, e.g., it is currently being deployed in Japan to transmit
3D content for TV broadcasting over the BS11 satellite
channel~\cite{JVC}.
The major drawback of these frame-compatible formats is that the
spatial sub-sampling requires interpolation
(and concomitant quality degradation) to
extract the left and right views at their original resolution.

\subsection{Multiview video compression}
\label{compres3d}
We now proceed to briefly introduce
the compression approaches that can be applied to the
representation formats outlined in the preceding subsection.
Building on the concept of inter-view
prediction~\cite{Lu}, multiview video coding~\cite{VeWS}
exploits the redundancies across different views of the same scene
(in addition to the temporal and intra-view spatial redundancies
exploited in single-view encoding).
Multiview video coding is applicable only to the
multiview (MV) representation format since this is the only format
to retain distinct sequences of video frames for the views.
For the case of 3D video, the recent official ITU
multiview video coding reference
software, referred to as JMVC, first encodes the left view, and then
predictively encodes the right view with respect to the encoded left view.

The frame sequential (FS) and side-by-side (SBS) representation
formats present a single sequence of video frames to the encoder.
Thus, conventional single-view video encoders can be applied to the FS
and SBS representations.
We employ the state-of-the-art JSVM reference
implementation~\cite{jsvm} of the scalable video coding (SVC) extension
of the advanced video coding (AVC) encoder in single-layer encoding mode.

For completeness, we briefly note that
each view could also be encoded independently with a
single-view encoder, which is referred to as simulcasting.
While simulcasting has the advantage of low
complexity, it does not exploit the redundancies between the views,
resulting in low encoding efficiency~\cite{VeTMC}.
A currently active research direction in multiview video encoding is
asymmetric coding~\cite{GuGST,SaGT11}, which
encodes the left and right views with different properties,
e.g., different quantization scales.
For other ongoing research directions in encoding, we refer
to the overviews in~\cite{VeTMC,GuGST,VeWS}.

\subsection{Multiview video streaming}
\paragraph{SBS representation} The $V = 2$ views are integrated into
one frame sequence with the spatial resolution and frame rate $f$ of the
underlying video.
For frame-by-frame transmission of a sequence with $M$ frames,
frame $m,\ m = 1, \ldots, M$, of size $X_m$ [bytes] is transmitted
during one frame period of duration $1/f$ at a bit rate of
$R_m = 8 f X_m$ [bit/s].

\paragraph{MV representation}
There are
a number of streaming options for the MV representation with $V$ views.
First, the $V$ streams resulting from the multiview video encoding
can be streamed individually.
We let $X_{m}(v),\ m = 1, \ldots, M,\ v = 1, \ldots, V$,
denote the size [bytes] of the encoded video frame $m$ of view $v$
and note that $R_m(v) = 8 f X_m(v)$ [bit/s] is the
corresponding bitrate.
The mean frame size of the encoded view $v$ is
\begin{equation}
  \label{eq:mean_frame_size}
  \bar{X}(v) = \frac{1}{M} \sum^{M}_{m=1}X_{m}(v)
\end{equation}
and the corresponding average bit rate is $\bar{R}(v) = 8 f \bar{X}(v)$.
The variance of these frame sizes is
\begin{equation}
  \label{eq:stddev}
  S^2_X(v) = \frac{1}{M - 1} \sum_{m=1}^{M}
              \left[ X_{m}(v)  - \bar{X}(v) \right]^2.
\end{equation}
The coefficient of variation of the frame sizes of view $v$ [unit free]
is the standard deviation of the frame sizes $S_X(v)$ normalized by the
mean frame size
\begin{equation}
  \label{eq:cov}
  CoV_X(v) = \frac{S_X(v)}{\bar{X}(v)}
\end{equation}
and is widely employed as a measure of the variability of the frame
sizes, i.e., the traffic bitrate variability.
Plotting the $CoV$ as a function of the quantization scale (or equivalently,
the average PSNR video quality) gives the bitrate variability-distortion (VD)
curve~\cite{paper46,ReLR02}.

Alternatively, the $V$ streams can be merged into one multiview stream.
We consider two elementary merging options, namely sequential (S) merging
and aggregation (combining).
With sequential merging, the $M$ frames of the $V$ views are temporally
multiplexed in round-robin fashion,
i.e., first view 1 of frame 1, followed by view 2 of frame 1,
$\ldots$, followed by view $V$ of frame 1, followed
by view 1 of frame 2, and so on.
From the perspective of the video transport system, each of these
$VM$ video frames (pictures) can be interpreted as a video frame to be
transmitted.
In this perspective,
the average frame size of the resulting multiview stream is
\begin{eqnarray}
\label{eq:avg1}
\bar{X} = \frac{1}{V} \sum_{v = 1}^V\bar{X}(v).
\end{eqnarray}
Noting that this multiview stream has $V$ frames to be played
back in each frame period of duration $1/f$,
the average bit rate of the multiview stream is
\begin{eqnarray} \label{barR:eqn}
\bar{R} = 8 V f \bar{X}.
\end{eqnarray}
The variance of the frame sizes of the
sequentially (S) merged multiview stream is
\begin{equation}
  \label{eq:stddev1}
  S^{2}_S = \frac{1}{(M - 1)(V-1)} \sum_{m,v=1}^{M,V}
           \left[ X_{m}(v) - \bar{X} \right]^2
\end{equation}
with the corresponding $\mathrm{CoV}_S =  S_S / \bar{X}$.

With combining (C), the $V$ encoded views corresponding
to a given frame index $m$ are aggregated to form one
\textit{multiview frame} of size $X_m = \sum_{v = 1}^V X_m(v)$.
For 3D video with $V = 2$, the pair of frames for a given
frame index $m$ (which corresponds to a given capture instant of the
frame pair) constitutes the multiview frame $m$.
Note that the average size of the multiview frames is
$V \bar{X}$ with $\bar{X}$ given in (\ref{eq:avg1}).
Further, note that these multiview frames have a rate of
$f$ multiview frames/s;
thus, the average bit rate of the multiview stream resulting from
aggregation is the same $\bar{R}$ as given in (\ref{barR:eqn}).
However, the variance of the sizes of the
(combined) multiview frames is different
from (\ref{eq:stddev1}); specifically,
\begin{equation}
  \label{eq:stddev2}
  S^{2}_C = \frac{1}{(M- 1)} \sum_{m=1}^{M}
           \left[ X_m  - V \bar{X} \right]^2
\end{equation}
and $\mathrm{CoV}_C =  S_C / (V \bar{X})$.

\paragraph{FS representation}
Similar to the MV representation, the FS representation
can be streamed sequentially (S) with the traffic characterizations given
by (\ref{eq:avg1})--(\ref{eq:stddev1}).
Or, the $V$ encoded frames for a given frame index $m$ can
be combined (C), analogous to the aggregation of the MV representation,
resulting in the frame size variance
given by (\ref{eq:stddev2}).

\paragraph{Frame size smoothing}
The aggregate streaming approach combines all encoded video data
for one frame period of playback duration $1/f$ [s] and transmits
this data at a constant bitrate over the $1/f$ period.
Compared to the sequential streaming approach,
the aggregate streaming approach thus performs smoothing
across the $V$ views, i.e., effectively smoothes
the encoded
video data over the duration of one frame period $1/f$.
This smoothing concept can be extended to multiple frame periods,
such as a Group of Pictures (GoP) of the encoder~\cite{AuRe09}.
For GoP smoothing with a GoP length of $G$ frames, the
encoded views from $G$ frames are aggregated and streamed at a constant
bitrate over the period $G/f$ [s].

\section{Evaluation Set-up}
\label{data}
In this section, we describe our evaluation set-up, including
the employed 3D video sequences, the encoding set-up, and the
video traffic and quality metrics.

\subsection{Video sequences}
\label{vid}
For a thorough evaluation of the traffic characteristics, especially
the traffic variability, the publicly available, relatively short
3D video research sequences~\cite{mic} are not well suited.
Therefore, we employ long 3D ($V = 2$) video sequences of
$M = 52,100$ frames each. That is, we employ 51,200 left-view
frames (pictures) and
51,200 right-view frames (pictures) for each test video.
We have conducted evaluations with
\textit{Monsters vs Aliens} and \textit{Clash of the Titans}, which are
computer-animated fiction movies,
\textit{Alice in Wonderland}, which is a fantasy movie consisting
of a mix of animated and real-life content, and
\textit{IMAX Space Station}, a documentary.
All videos are in the full HD $1920\times 1080$ pixels format and have
a frame rate of $f = 24$ frames/s for each view.

\subsection{Encoding set-up}
We encoded the multiview representation with the reference software JMVC
(version 8.3.1).
We encoded the FS and SBS representations with
the broadly used H.264 SVC video coding standard using the
H.264 reference software JSVM (version 9.19.10)~\cite{jsvm,VaDR08}
in single-layer encoding mode.
We set the GOP length to $G = 16$ frames for the MV and SBS encodings;
for the FS encodings we set $G = 32$ so that all encodings have the
same playback duration between intracoded (I) frames, i.e., support the same
random access granularity.
We employ two different GoP patterns:
(B1) with one bi-directionally predicted (B) frame between
successive intracoded (I) and predictive encoded (P) frames, 
and (B7) with seven B frames between successive I and P frames.
We conducted encodings for the quantization parameter settings
24, 28, and 34.

\subsection{Traffic and quality metrics}

We employ the peak signal to noise ratio (PSNR)~\cite{ChSRK11,SeSBC10}
of the video luminance signal of a video frame
$m,\ m = 1, \ldots, M$, of a view $v,\ v = 1,\ldots, V$,
as the objective quality metric of
video frame $m$ of view $v$. We average these video frame PSNR values
over the $VM$ video frames of a given video sequence
to obtain the average PSNR video quality.
For the MV and FS representations, the PSNR evaluation is conducted
over the full HD spatial resolution of each view of a given frame.
We note that in the context of
asymmetric 3D video coding~\cite{SaGT11},
the PSNR values of the two views may be weighed unequally, depending
on their relative scaling (bit rate reduction)~\cite{OzTT09}.
We  do not consider asymmetric video coding
in this study and weigh the PSNR values of
both views equally.

For the SBS representation, we report 
for some encodings the PSNR values from the comparison
of the uncompressed SBS representation with the encoded (compressed)
and subsequently decoded SBS representation
as SBS without interpolation (SBS-NI).
We also report for all encodings 
the comparison of the original full HD left and right views
with the video signal obtained after SBS representation, encoding,
decoding, and
subsequent interpolation back to full HD format as
SBS with interpolation (SBS-I).
Unless otherwise noted, the SBS results are for the SBS representation
with interpolation.
We employed the JSVM
reference down-sampling with a Sine-windowed
Sinc-function and the corresponding normative up-sampling
using a set of 4-tap filters~\cite{jsvm}.
We plot the average
PSNR video quality [dB] as a function of the average streaming
bitrate $\bar{R}$ [bit/s] to obtain the RD curve and the
coefficient of variation of the frame sizes $\mathrm{CoV}$
as a function of the average PSNR video quality to obtain the VD curve.

\section{Traffic and Quality Results}
\label{reslt}
In this section we present the RD and VD characteristics for the
examined 3D video representation formats.
We briefly note that generally the encodings with one B frame between
successive I and P frames follow the same trends as observed for the
encodings with seven B frames; the main difference is that the encodings
with one B frame have slightly higher bitrates and slightly lower CoV
values, which are effects of the lower level of predictive encoding.

\label{res}
\begin{figure*}[!htb]
 \begin{tabular}{cc}
  \includegraphics[height=.34\textheight,angle=270]{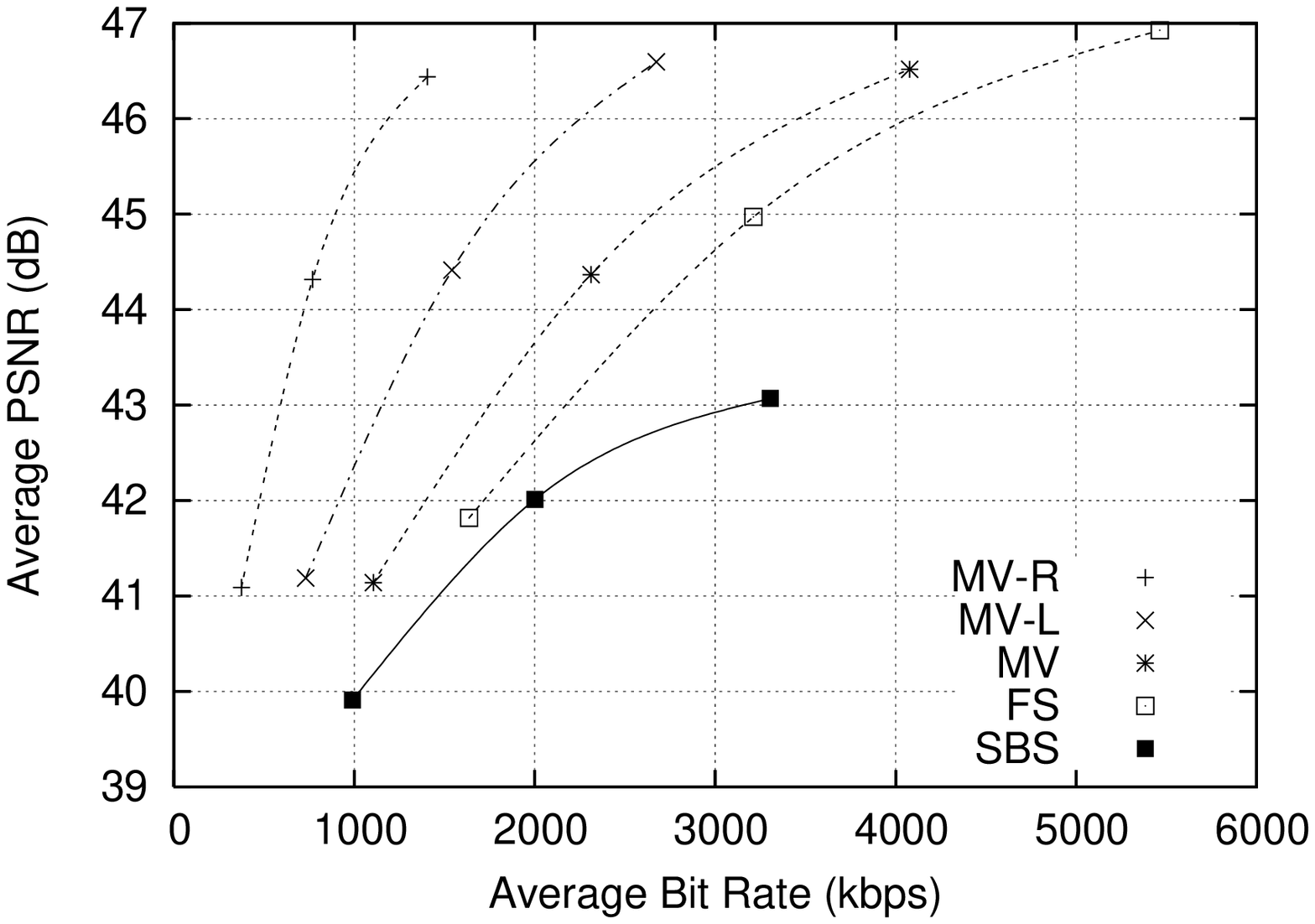} &
    \includegraphics[height=.34\textheight,angle=270]{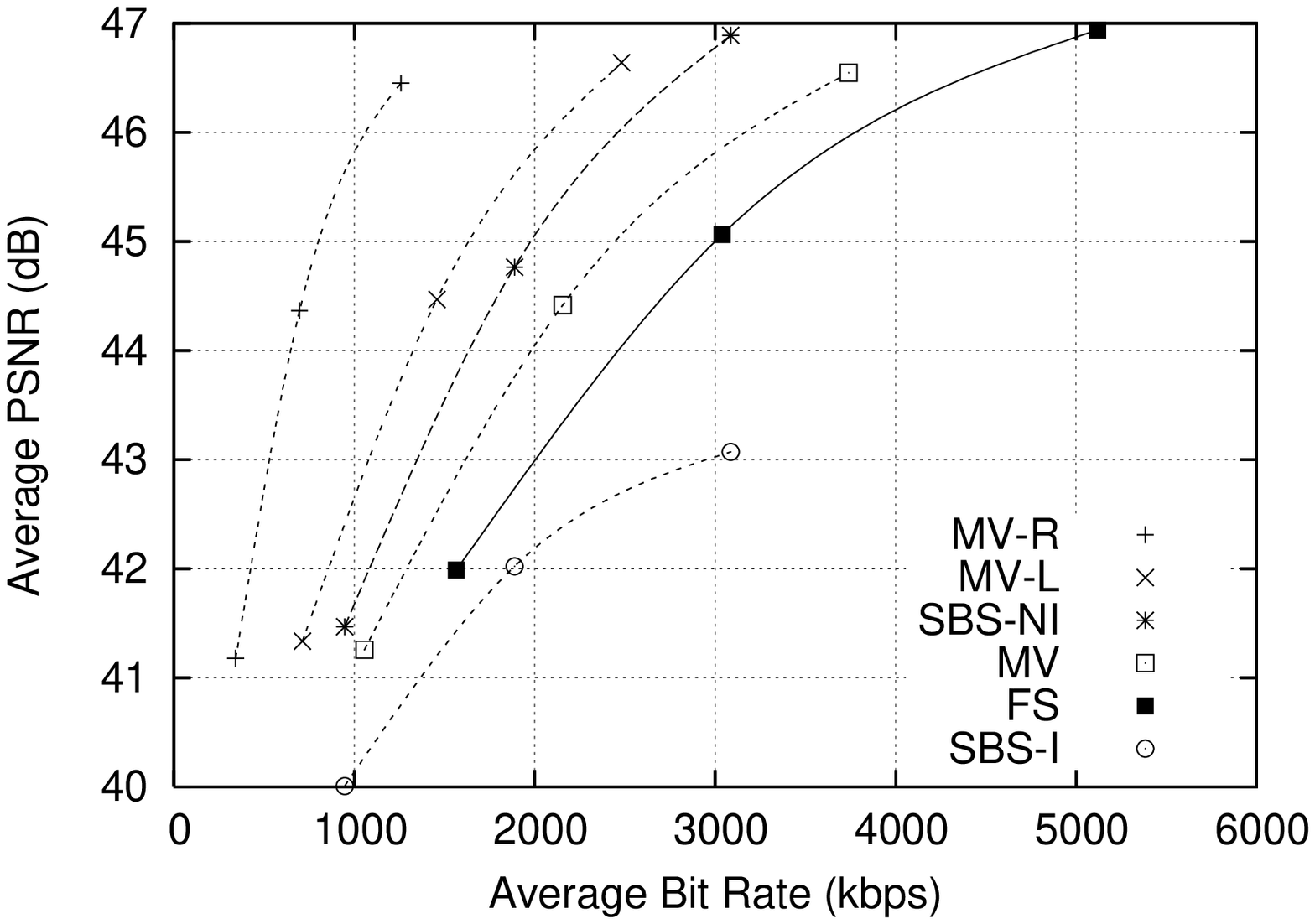}         \\
  {\footnotesize (a) \textit{Monsters vs Aliens}, B1}  &
    {\footnotesize (b) \textit{Monsters vs Aliens}, B7}   \\
     \includegraphics[height=.34\textheight,angle=270]{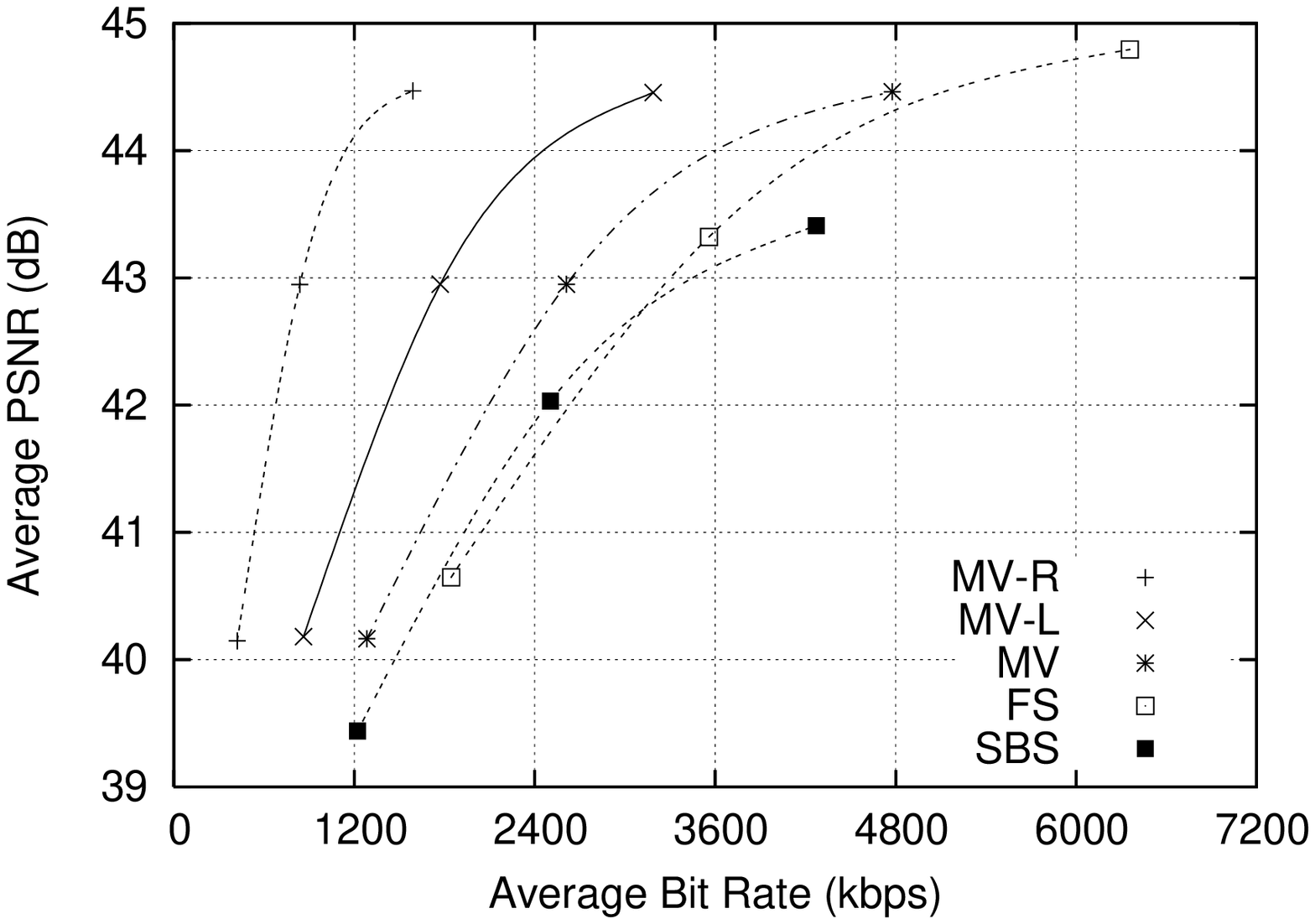} &
       \includegraphics[height=.34\textheight,angle=270]{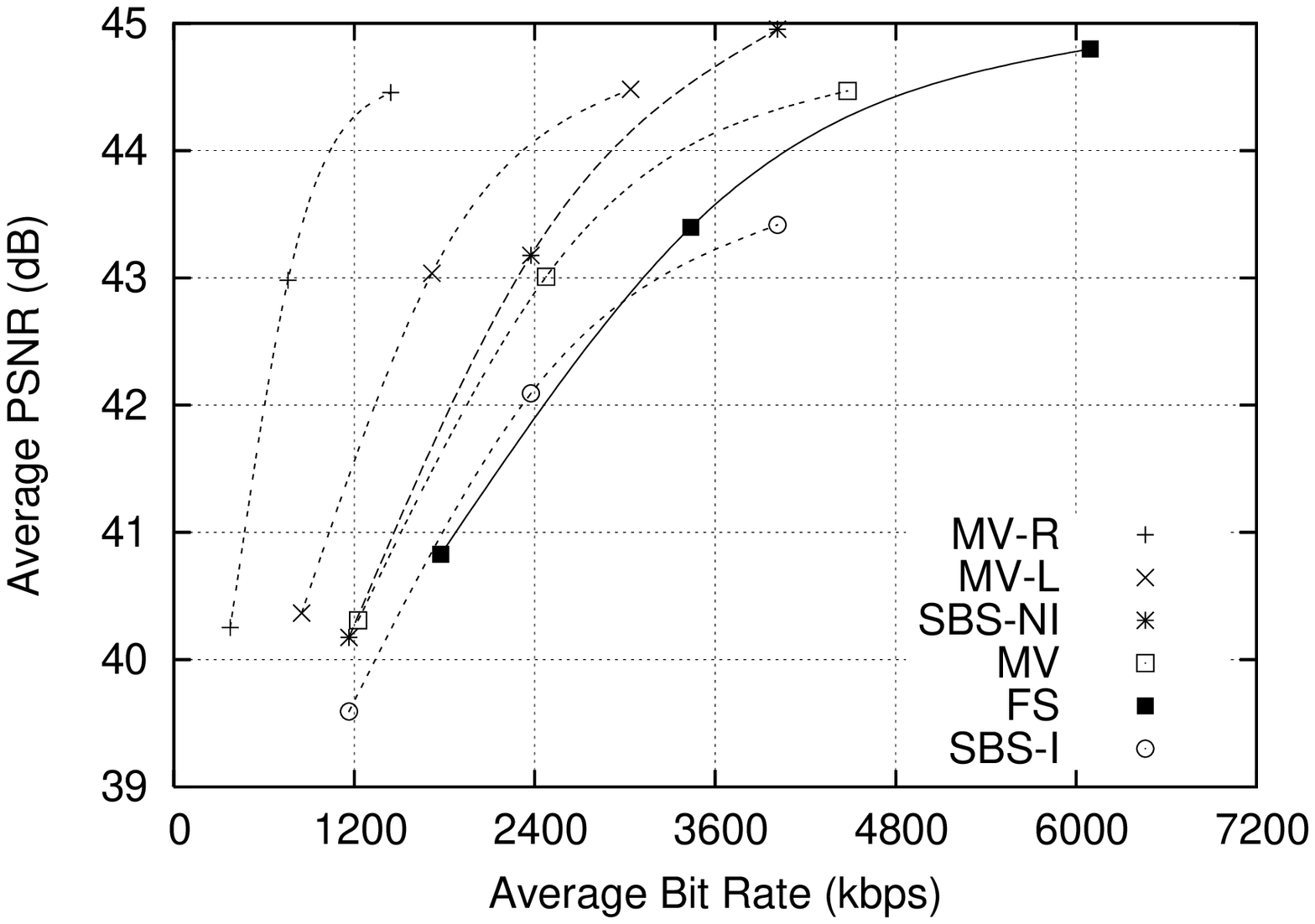}     \\
   {\footnotesize (c) \textit{Alice in Wonderland}, B1}  &
      {\footnotesize (d) \textit{Alice in Wonderland}, B7}    \\
      \includegraphics[height=.34\textheight,angle=270]{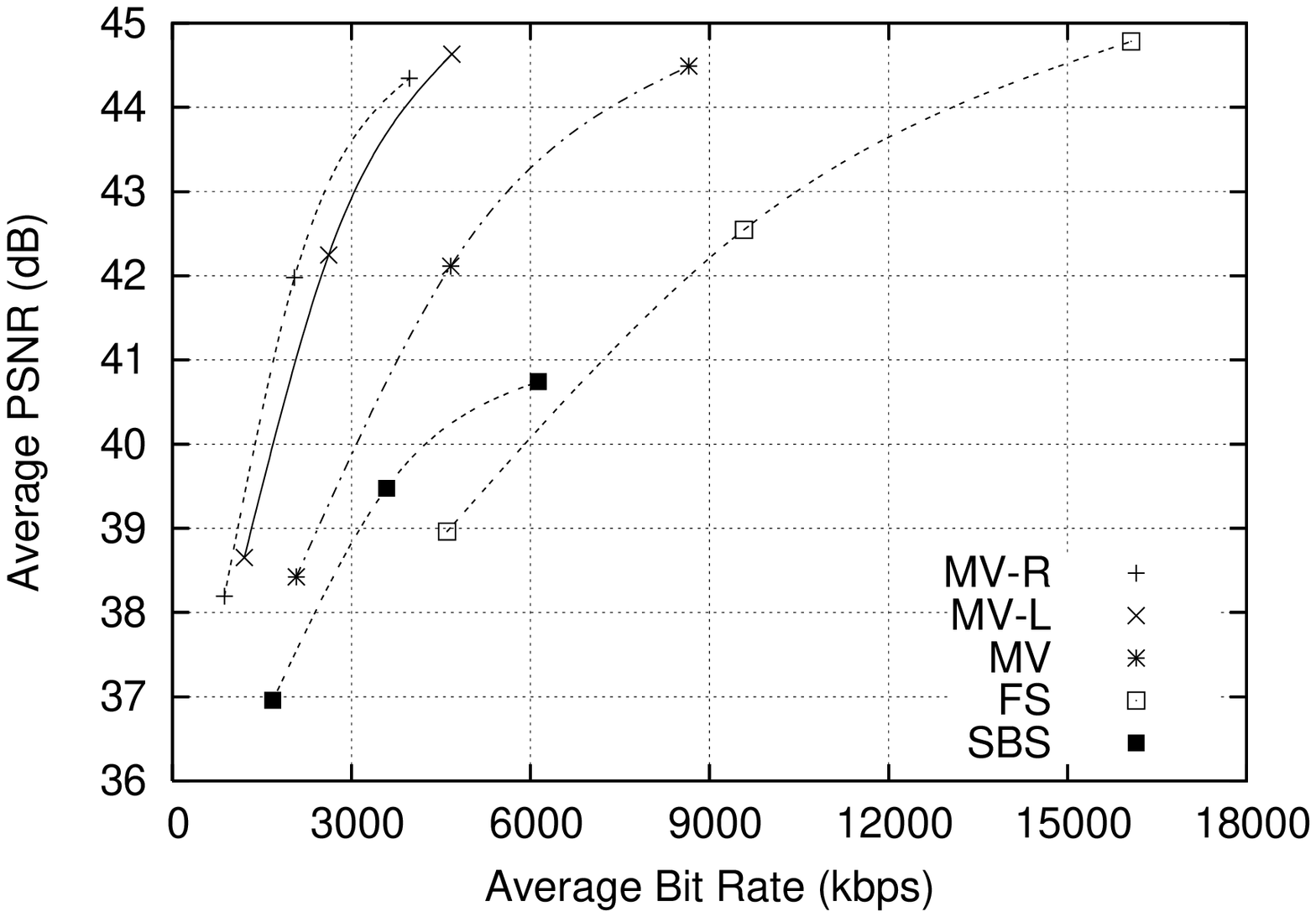} &
      \includegraphics[height=.34\textheight,angle=270]{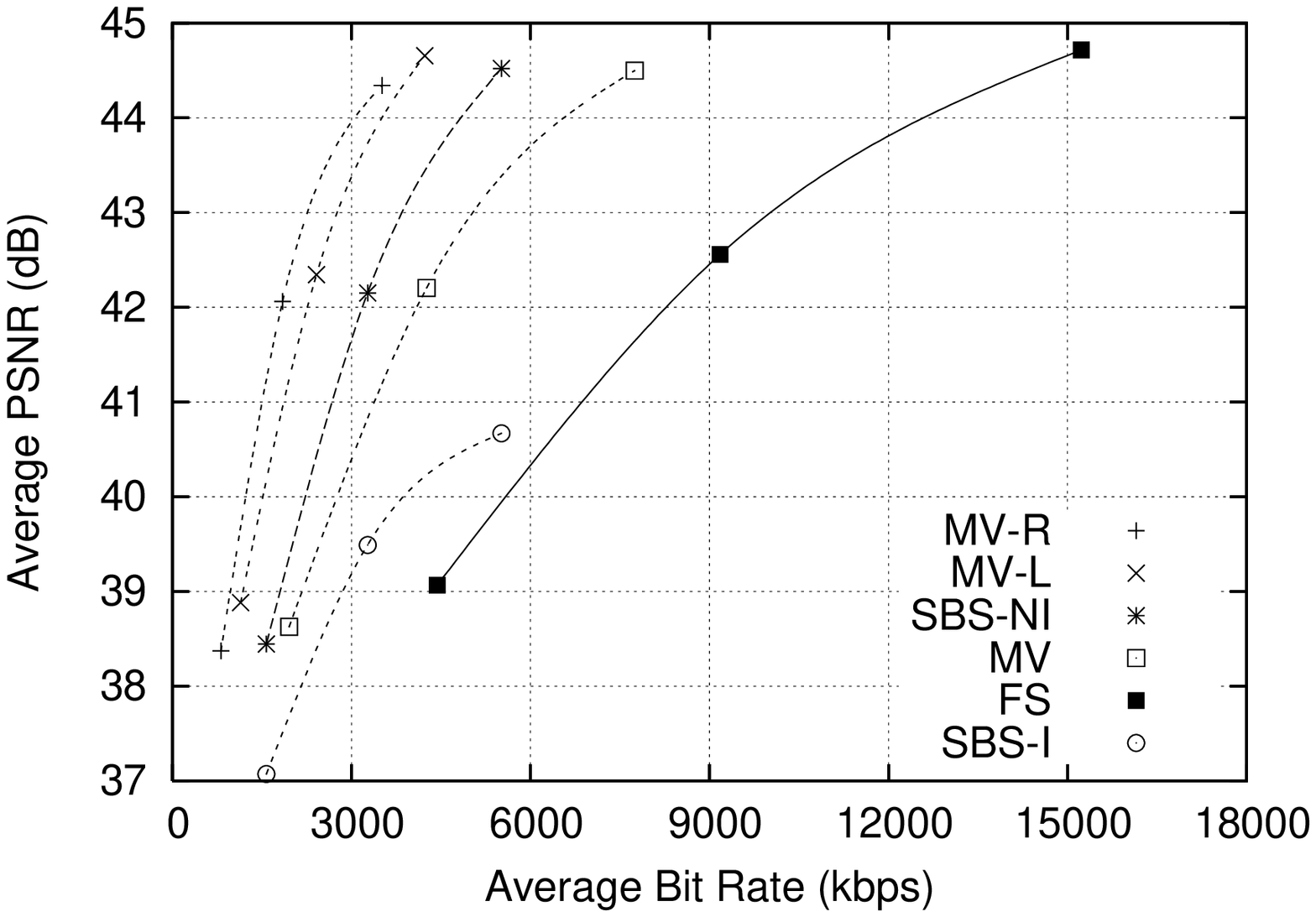}  \\
  {\footnotesize (e) \textit{IMAX Space Station}, B1}  &
   {\footnotesize (f) \textit{IMAX Space Station}, B7} \\  
\end{tabular}
    \caption{RD curves for multiview (MV) representation,
    frame sequential (FS) representation, and
    side-by-side (SBS) representation for GoP patterns with
   one B frame between successive I and P frames (B1)
   as well as seven B frames between successive I and P frames (B7).}
    \label{rdall}
\end{figure*}
\subsection{Bitrate-distortion (RD) characteristics}
\label{rd}
In Fig.~\ref{rdall}, we plot the RD curves of
the multiview representation encoded with the multiview video codec
for streaming the left view only (MV-L), the right view only (MV-R),
and the merged multiview stream (MV).
Similarly, we plot the RD curves for the frame sequential (FS) representation
and the side-by-side (SBS) representation
encoded with the conventional single-view codec.

From the MV curves in Fig.~\ref{rdall}, we observe that the right
view has a significant RD improvement compared to the left view. This is
because of the inter-view prediction of the multiview encoding, which
exploits the inter-view
redundancies by encoding the right view with prediction from the left view.

Next, turning to the side-by-side (SBS) representation, we observe
that SBS with interpolation can
achieve similar or even slightly better RD efficiency than FS 
for the low to medium quality range of
videos with real-life content (\textit{Alice in Wonderland} and
\textit{IMAX Space Station}).
However, SBS has
consistently lower RD efficiency than the MV representation.
In additional evaluations for the B7 GoP pattern, 
we compared the uncompressed SBS
representation with the encoded (compressed) and subsequently
decoded SBS representation and found that the RD curve for this
SBS representation without interpolation (SBS-NI) lies between the 
MV-L and MV RD curves.
We observed from these additional evaluations 
that the interpolation to the full HD format (SBS-I) significantly reduces the
average PSNR video quality, 
especially for encodings in the higher quality range.

Finally, we observe from Fig.~\ref{rdall} that the MV representation
in conjunction with multiview encoding has
consistently higher RD efficiency than the FS representation with
conventional single-view encoding.
The FS representation essentially translates the
multi-view encoding problem into a temporally predictive coding problem.
That is, the FS representation temporally interleaves the left and right
views and then employs state-of-the-art temporal predictive encoding.
The results in Fig.~\ref{rdall} indicate that this temporal predictive
encoding can not exploit the inter-view redundancies as well as the
state-of-the-art multiview encoder.

\begin{figure*}[!htb]
 \begin{tabular}{cc}
      \includegraphics[height=.34\textheight,angle=270]{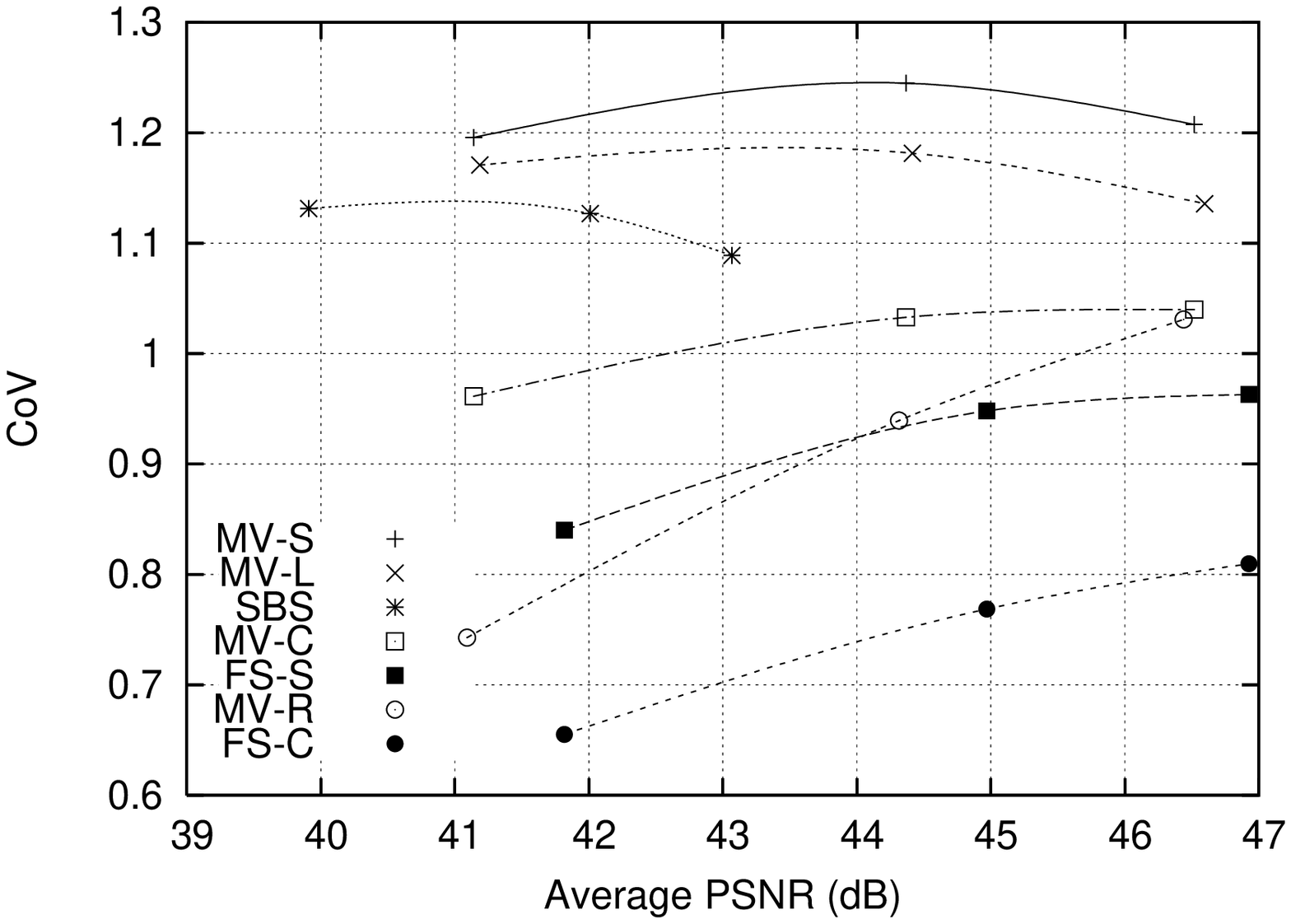} &
      \includegraphics[height=.34\textheight,angle=270]{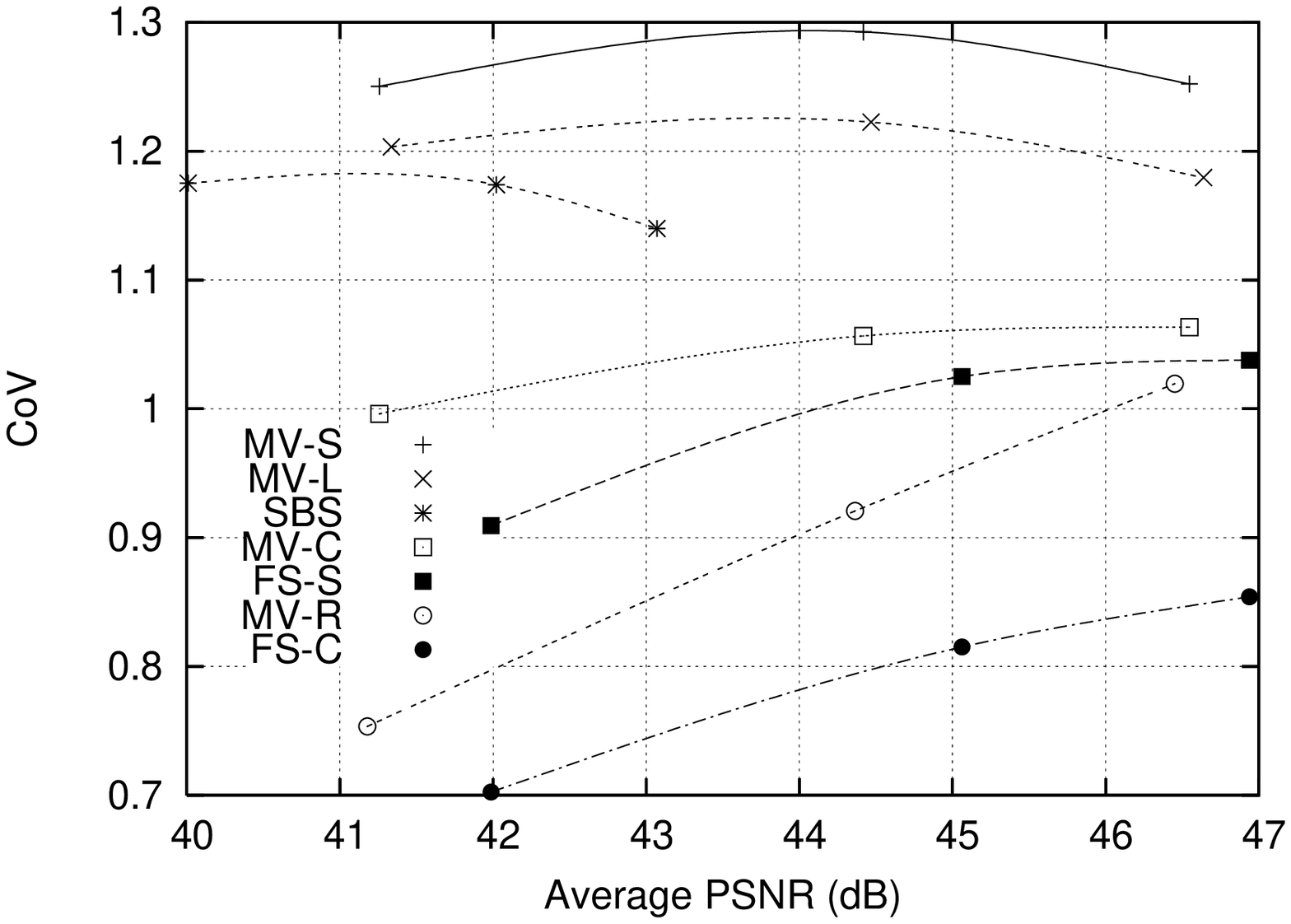} \\
    {\footnotesize (a) \textit{Monsters vs Aliens}, B1}  &
     {\footnotesize (b) \textit{Monsters vs Aliens}, B7}      \\
      \includegraphics[height=.34\textheight,angle=270]{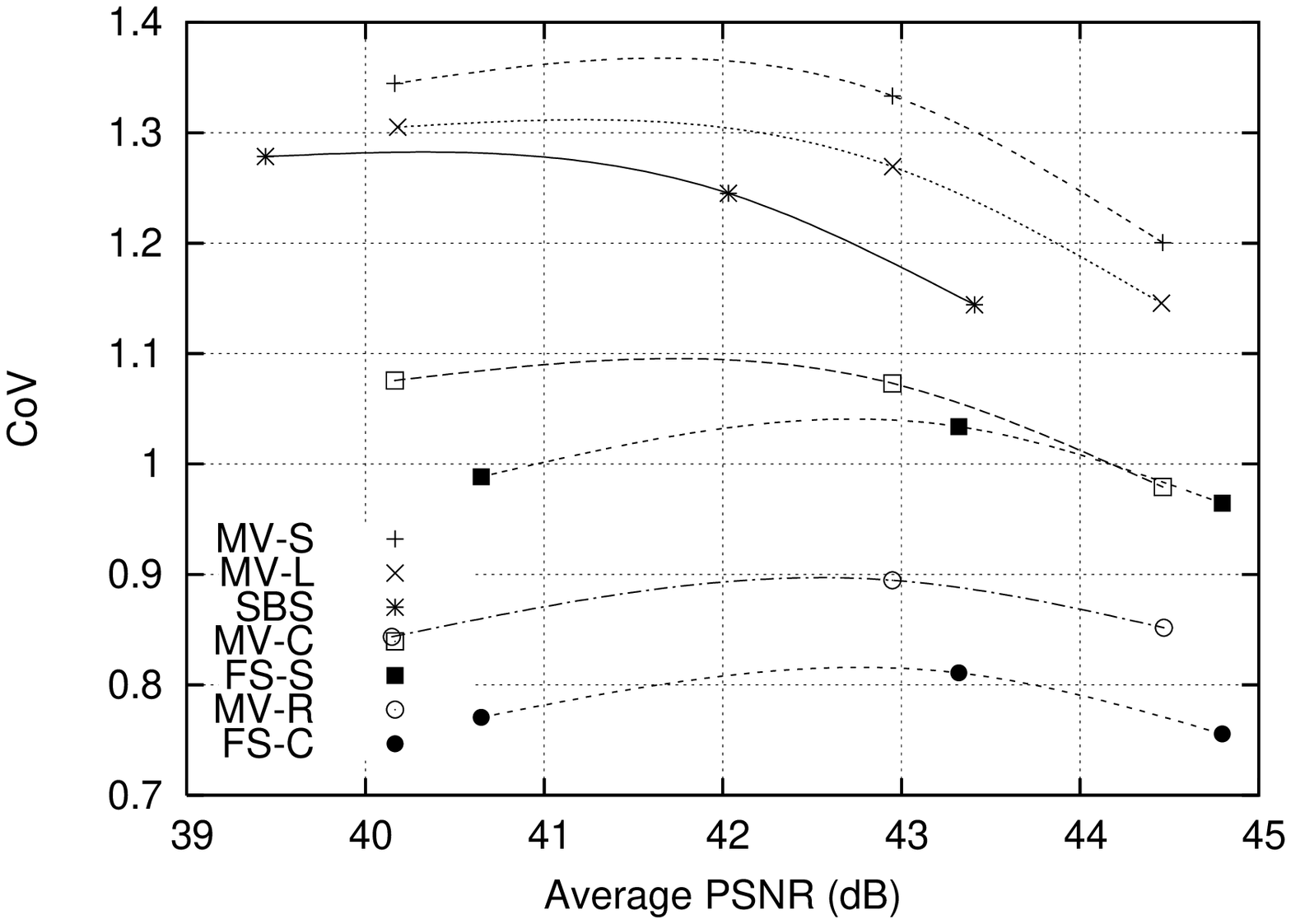} &
      \includegraphics[height=.34\textheight,angle=270]{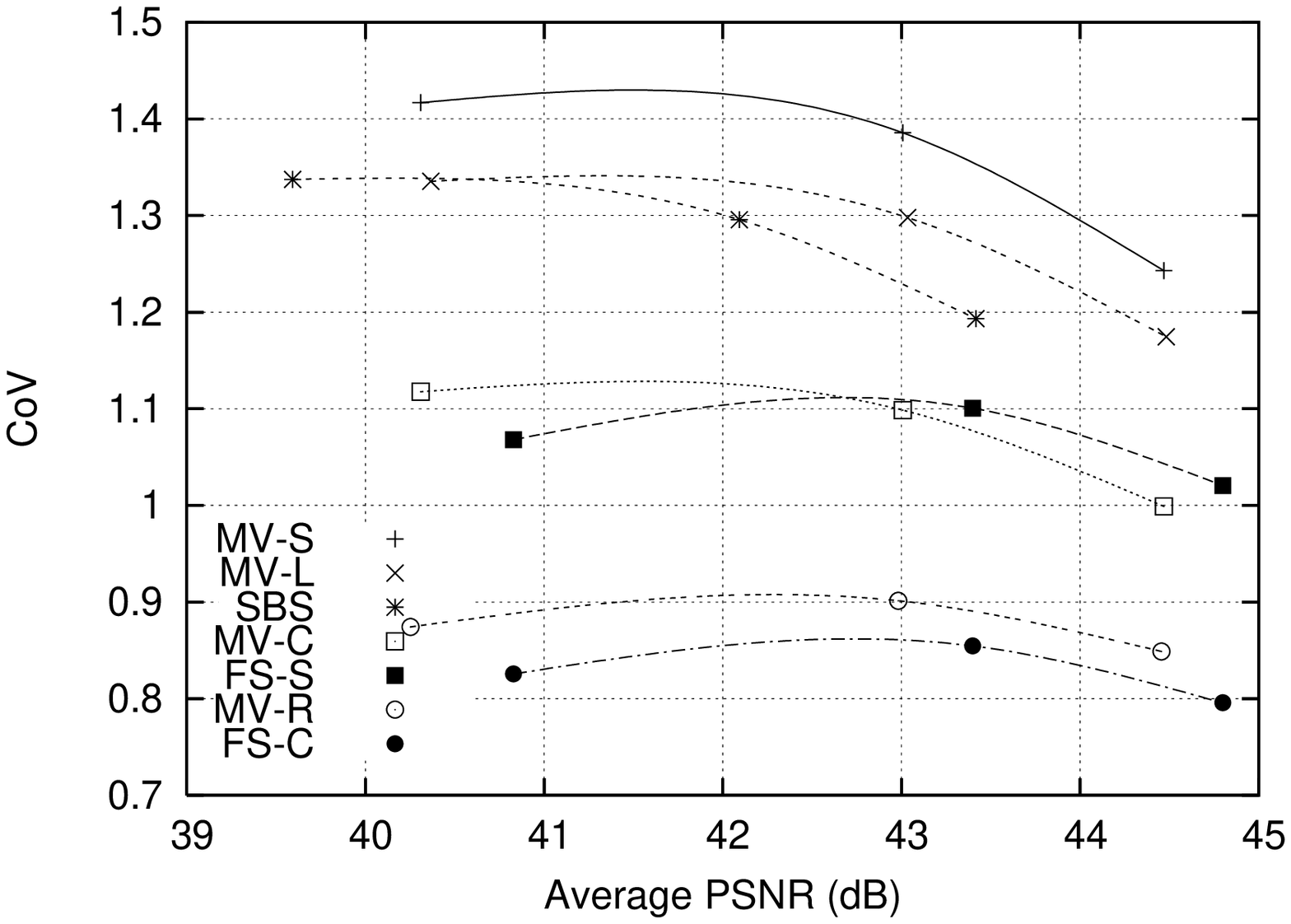}      \\
           {\footnotesize (c) \textit{Alice in Wonderland}, B1} &
           {\footnotesize (d) \textit{Alice in Wonderland}, B7}     \\
      \includegraphics[height=.34\textheight,angle=270]{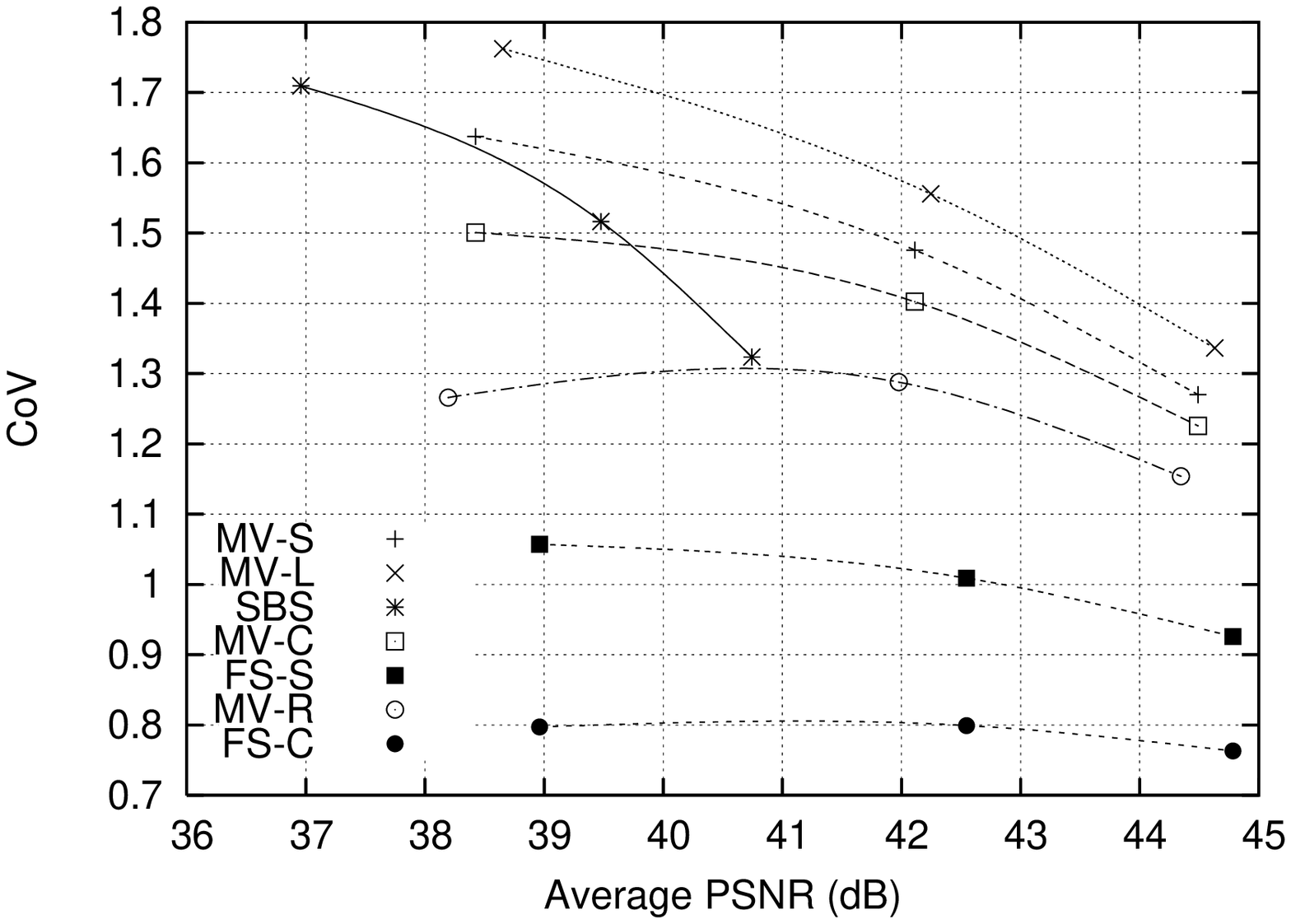} &
        \includegraphics[height=.34\textheight,angle=270]{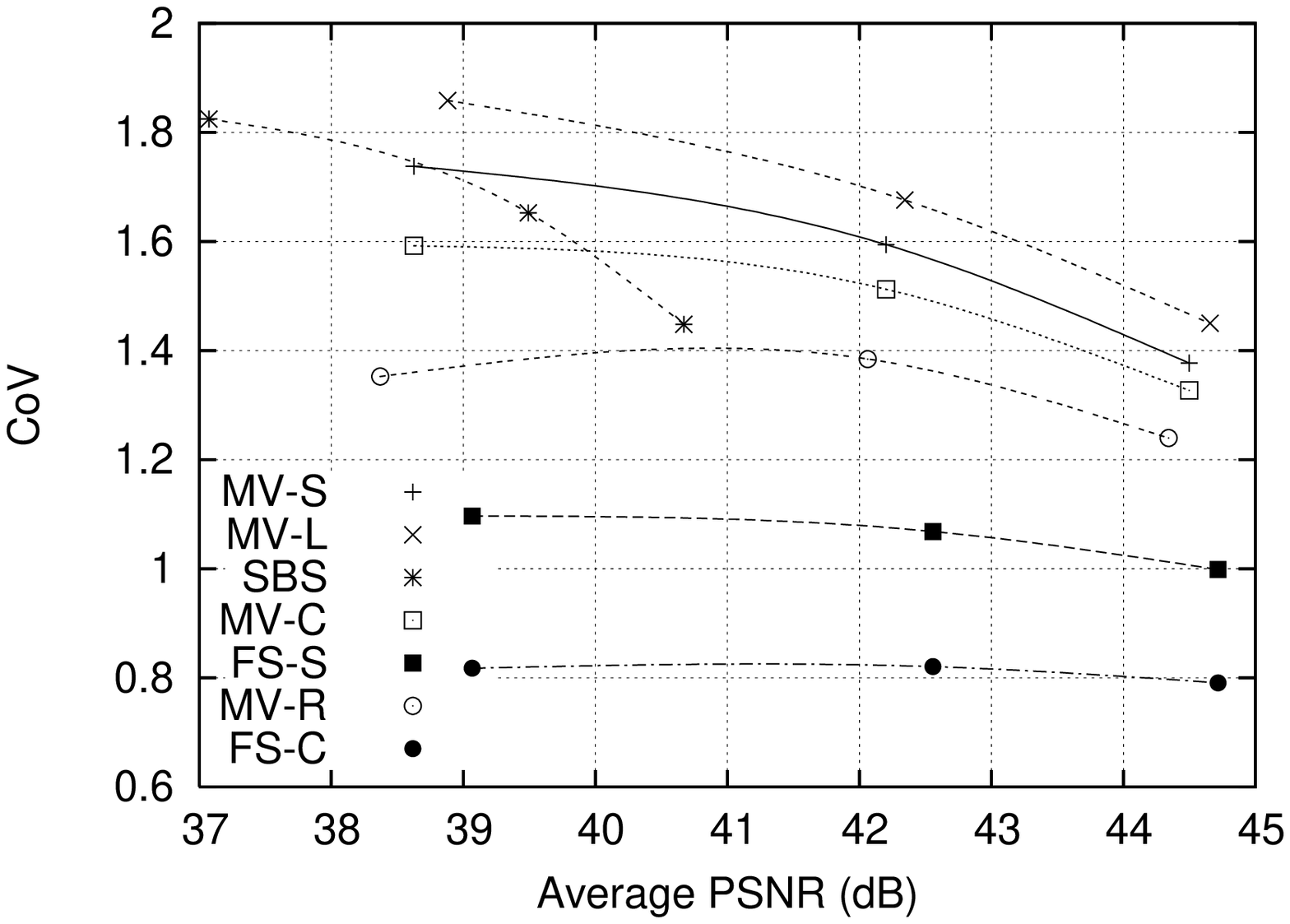}     \\
           {\footnotesize (e) \textit{IMAX Space Station}, B1} &
             {\footnotesize (f) \textit{IMAX Space Station}, B7} \\       \\
\end{tabular}
    \caption{VD curves for different representation formats
  and streaming approaches for B1 and B7 GoP patterns.}
    \label{vdall}
\end{figure*}
\subsection{Bitrate variability-distortion (VD) characteristics}
\label{VD}
In Fig.~\ref{vdall}, we plot the VD curves for the examined
multiview (MV), frame sequential (FS), and side-by-side (SBS) representation
formats; whereby, for MV and FS, we plot both VD curves for
sequential (S) merging and aggregation (C).
We first observe from Fig.~\ref{vdall} that the MV representation
with sequential streaming (MV-S) has the highest traffic variability.
This high traffic variability is primarily due to the
size differences between successive encoded left and right views.
In particular, the left view is encoded independently and is thus
typically large.
The right view is encoded predictively from the left view and thus
typically small.
This succession of large and small views (frames), whereby each view is
treated as an independent video frame by the transmission system, i.e., is
transmitted within half a frame period $1/(2f)$ in the sequential
streaming approach, leads to the high traffic variability.
Smoothing over one frame period $1/f$ by combing the two views of each frame
from the MV encoding significantly reduces the traffic variability.
In particular, the MV encoding with aggregation (MV-C) has
generally lower traffic variability than the SBS streams.

We further observe from the MV results in Fig.~\ref{vdall} that the
left view (MV-L)
has significantly higher traffic variability than the right view (MV-R).
The large MV-L traffic variabilities are primarily due to the typically
large temporal variations in the scene content of the videos, which result
in large size variations of the MV-L frames which are encoded with temporal
prediction across the frames of the left view.
In contrast, the right view is predictively encoded from the left view.
Due to the marginal difference between the two perspectives of
the scene employed for the two views of 3D video, the content variations
between the two views (for a given fixed frame index $m$) are small
relative to the scene content variations occurring over time.

Turning to the FS representation, we observe that FS with sequential streaming
has CoV values near or below the MV representation with aggregation.
Similarly to the MV representation, aggregation significantly reduces the
traffic variability of the FS representation.
In fact, we observe from Fig.~\ref{vdall} that the FS representation
with aggregation has consistently the lowest CoV values.
The lower traffic variability of the FS representation is
consistent with its relatively less RD-efficient encoding.
The MV representation and encoding exploits the inter-view
redundancies and thus encodes the two views of each frame
more efficiently,
leading to relatively larger variations in the encoded frame sizes
as the video content and scenes change and present varying
levels of inter-view redundancy.
The FS representation with single-view encoding, on the other hand,
is not able to exploit these varying degrees of inter-view redundancy
as well,
resulting in less variability in the view and frame sizes, but also
larger average frame sizes.

In additional evaluations that are not included here in detail due
to space constraints, we found that frame size smoothing over
one GoP reduces the traffic variability significantly,
especially for the burstier MV representation.
For instance, for \textit{Monsters vs Aliens},
the CoV value of 1.05 for MV-C
for the middle point in Fig.~\ref{vdall}(a) is reduced
to 0.65 with GoP smoothing. Similarly, the corresponding CoV value of
1.51 for \textit{IMAX Space Station} (Fig.~\ref{vdall}(c)) is reduced to 0.77.
The CoV reductions are less pronounced for the FS representation:
the middle CoV value of 0.81 for FS-C in Fig.~\ref{vdall}(a)
is reduced to 0.58, while the corresponding CoV value of 0.82 in
Fig.~\ref{vdall}(c) is reduced to 0.70.

\section{Statistical Multiplexing Evaluations}
\label{impl}
In this section, we conduct statistical multiplexing
simulations to examine the impact of the 3D video representations on
the bandwidth requirements for streaming with minuscule
loss probabilities~\cite{RoMV96}.
For the MV and FS representations,
we consider the combined (C) streaming approach
where the pair of frames for each frame index $m$ is aggregated and
the GoP smoothing approach.
\begin{figure*}[!htb]
 \begin{tabular}{cc}
  \includegraphics[height=.34\textheight,angle=270]{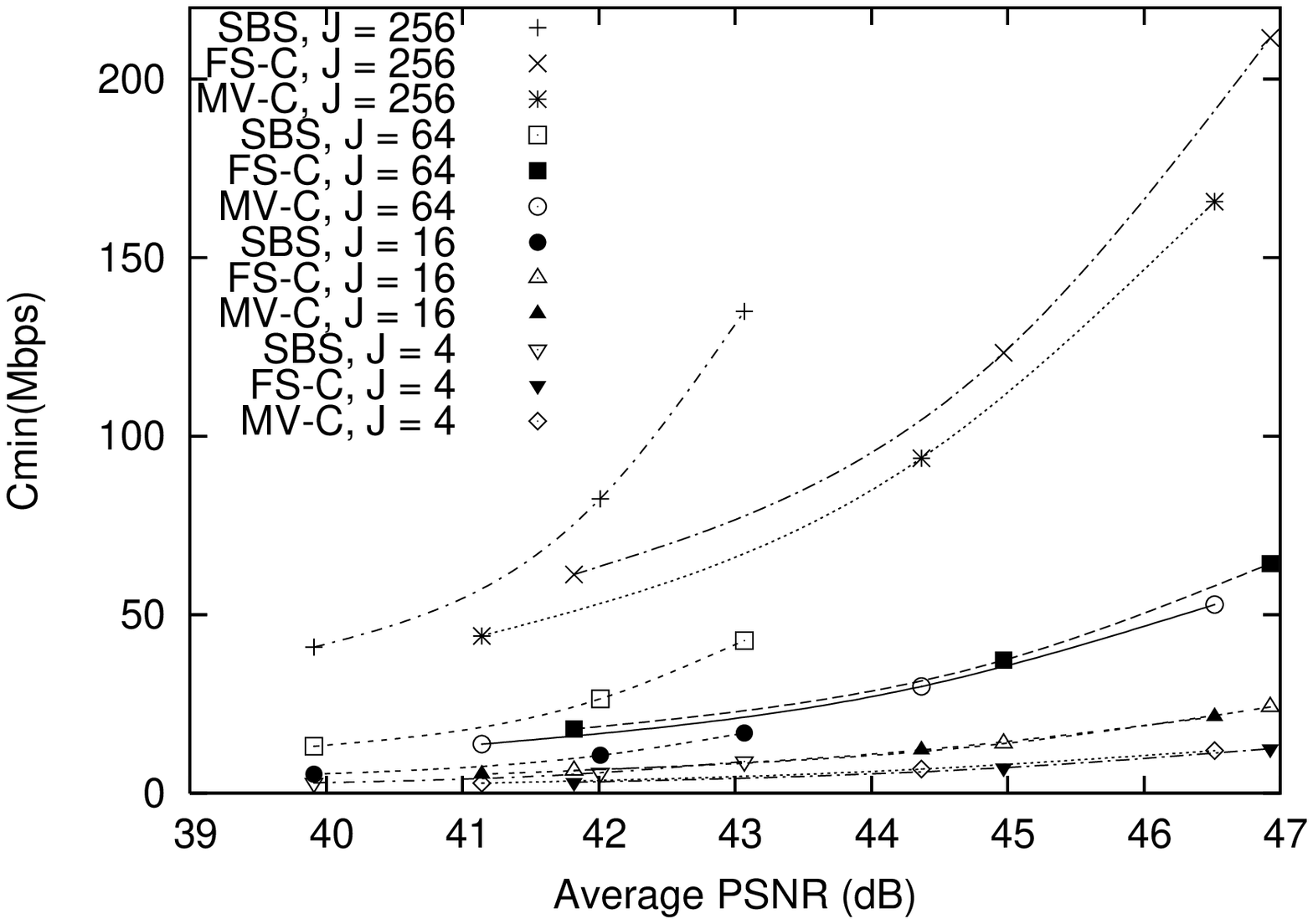} &
  \includegraphics[height=.34\textheight,angle=270]{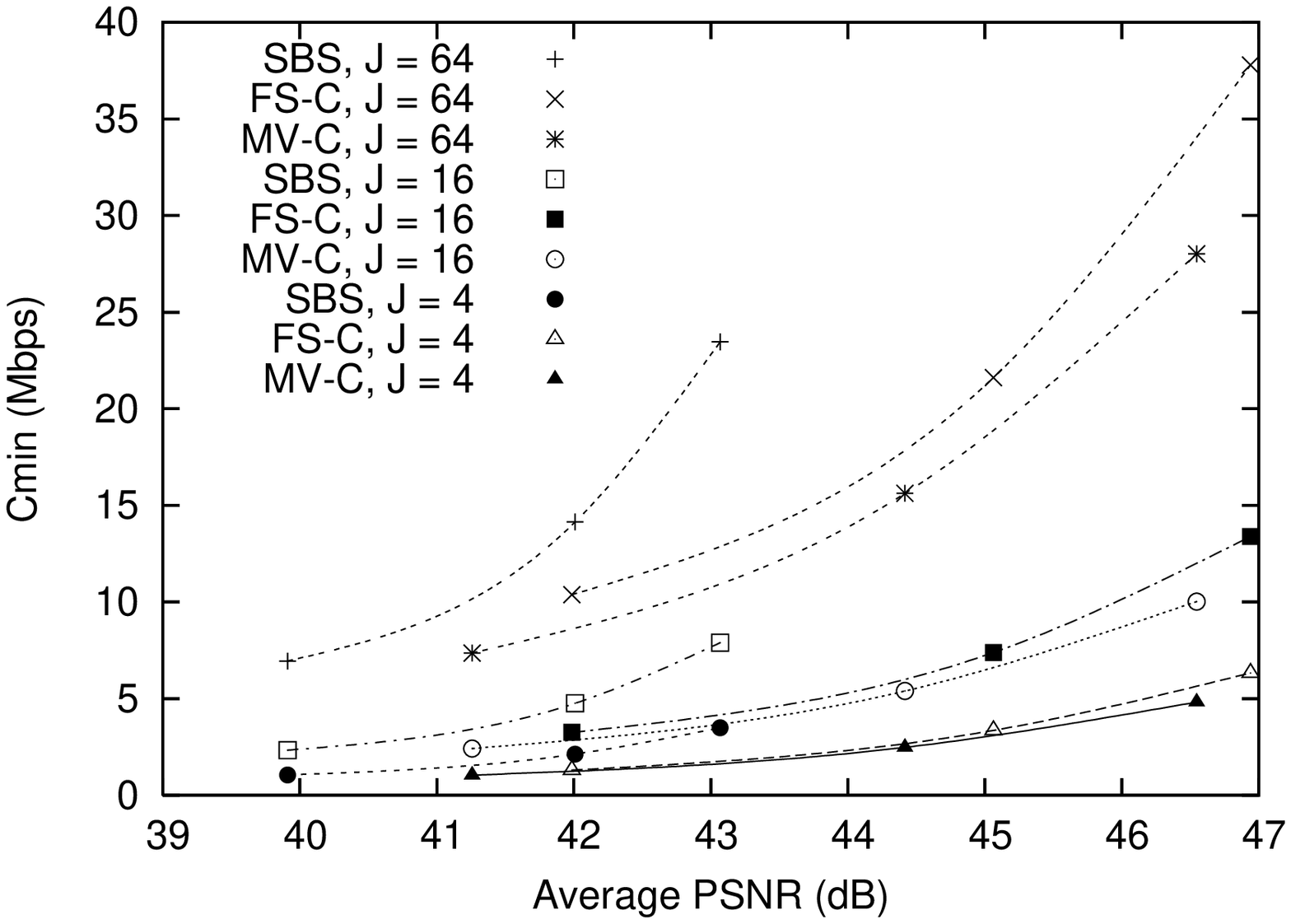} \\
  {\footnotesize (a) \textit{Monsters vs Aliens}, frame-by-frame} &
  {\footnotesize (b) \textit{Monsters vs Aliens}, GoP smoothing} \\
      \includegraphics[height=.34\textheight,angle=270]{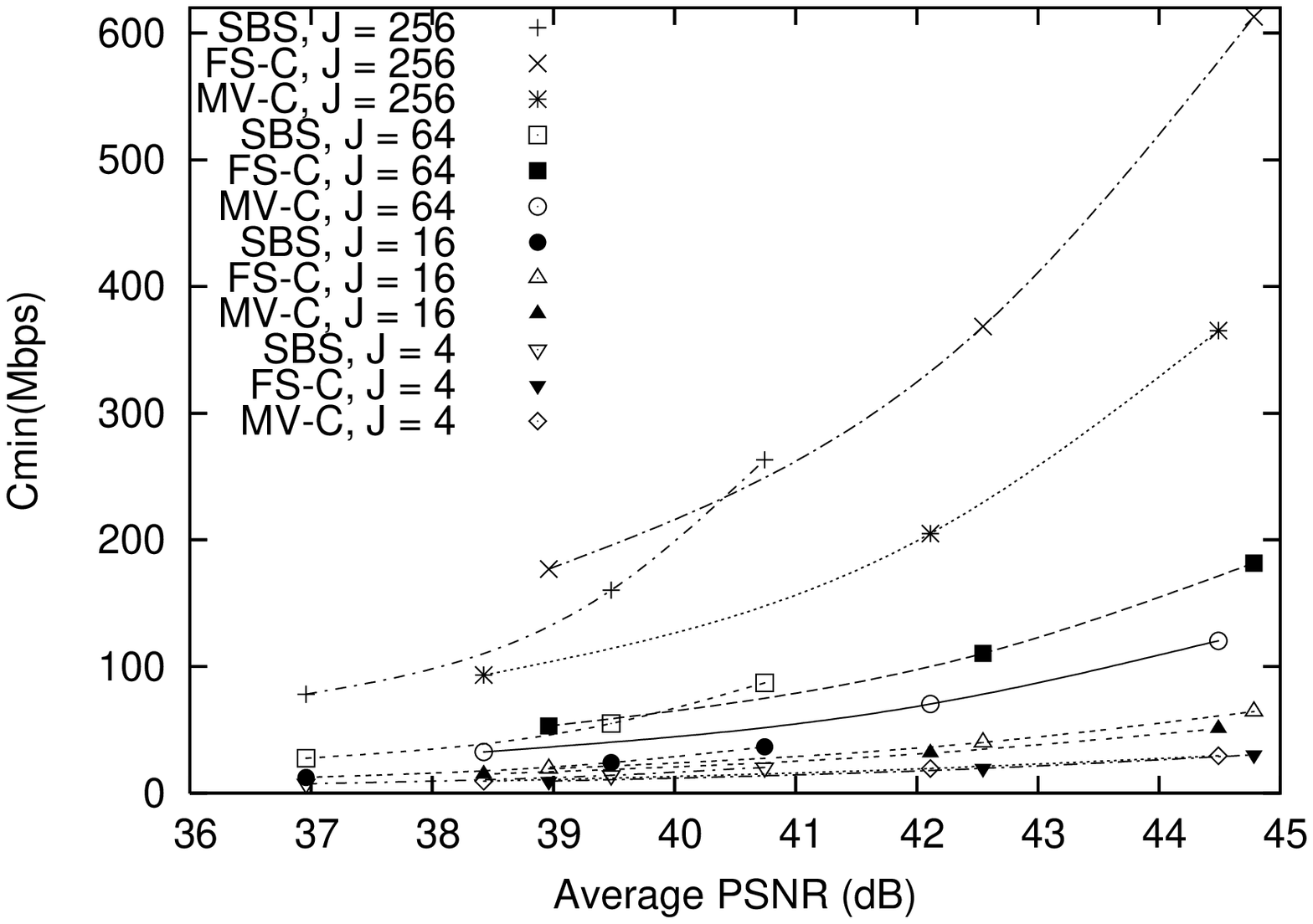} &
      \includegraphics[height=.34\textheight,angle=270]{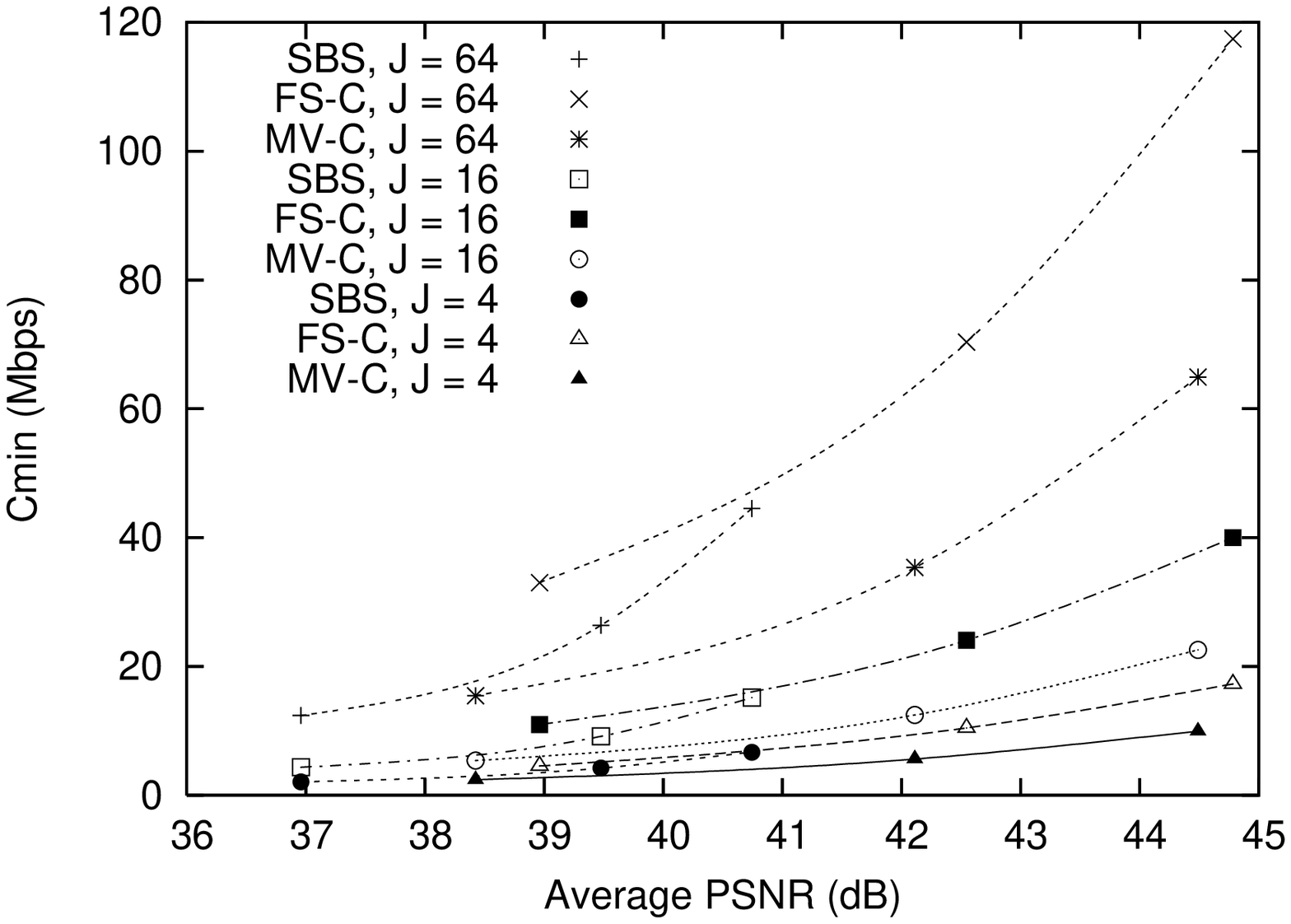} \\
           {\footnotesize (c) \textit{IMAX Space Station}, frame-by-frame} &
           {\footnotesize (d) \textit{IMAX Space Station}, GoP smoothing} \\
\end{tabular}
    \caption{Required minimum link transmission bit rate $C_{\min}$
    to transmit $J$ streams with an information loss probability
   $P_{\rm loss}^{\rm info} \leq \epsilon = 10^{-5}$.
    GoP structure B1 with one B frame between I and P frames.}
    \label{CminallB1}
\end{figure*}
\begin{figure*}[!htb]
 \begin{tabular}{cc}
  \includegraphics[height=.34\textheight,angle=270]{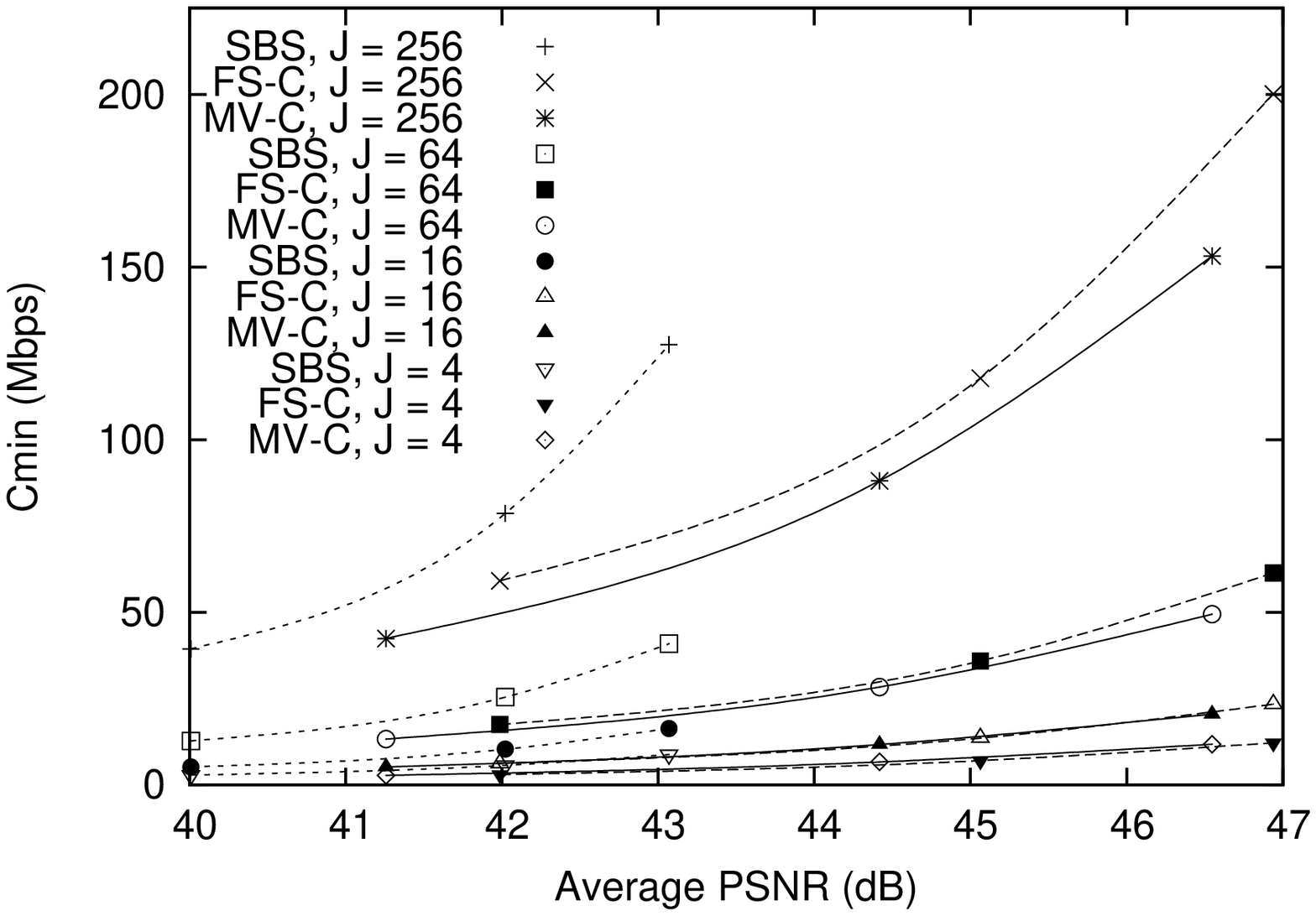} &
  \includegraphics[height=.34\textheight,angle=270]{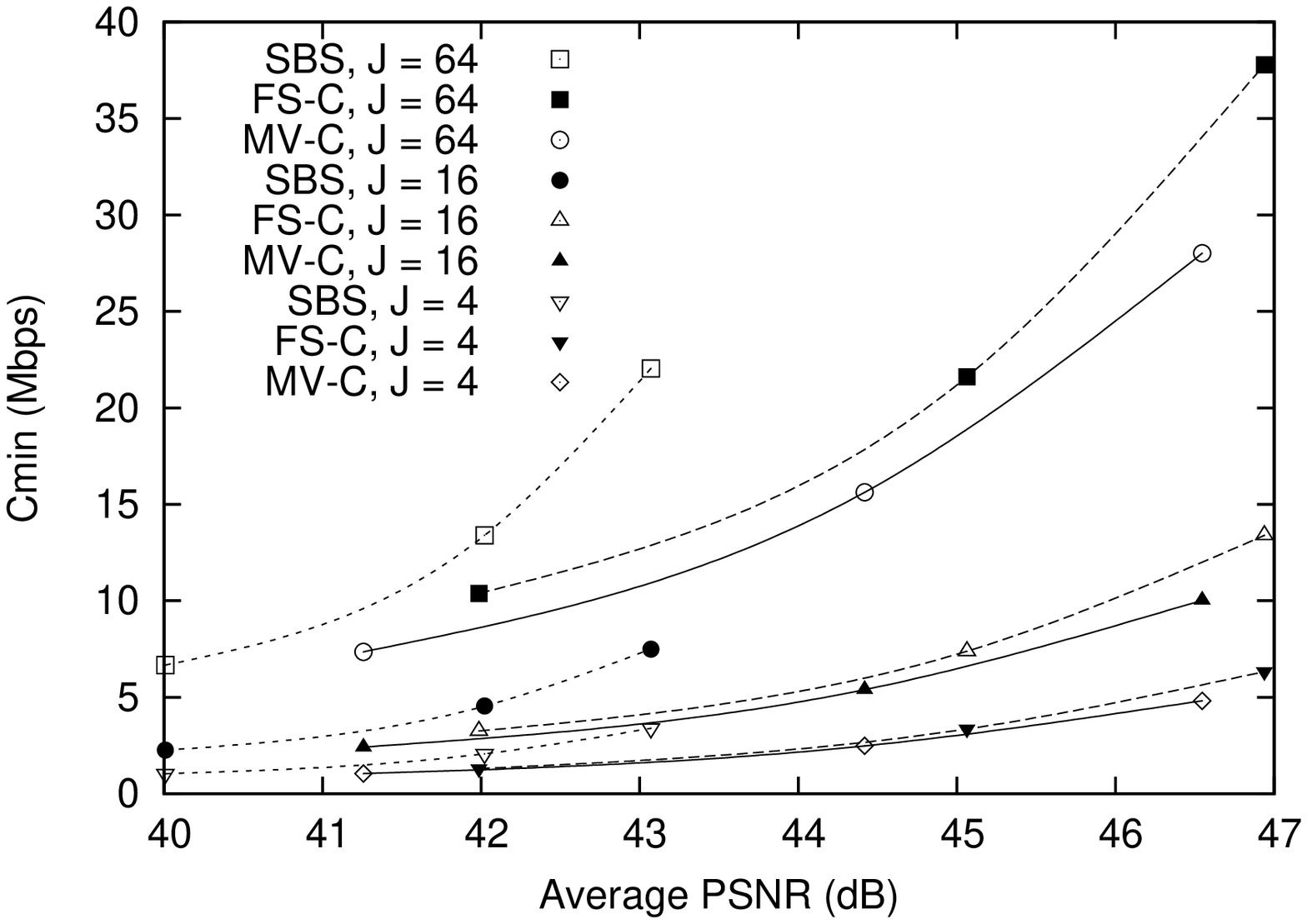} \\
  {\footnotesize (a) \textit{Monsters vs Aliens}, frame-by-frame} &
  {\footnotesize (b) \textit{Monsters vs Aliens}, GoP smoothing} \\
  \includegraphics[height=.34\textheight,angle=270]{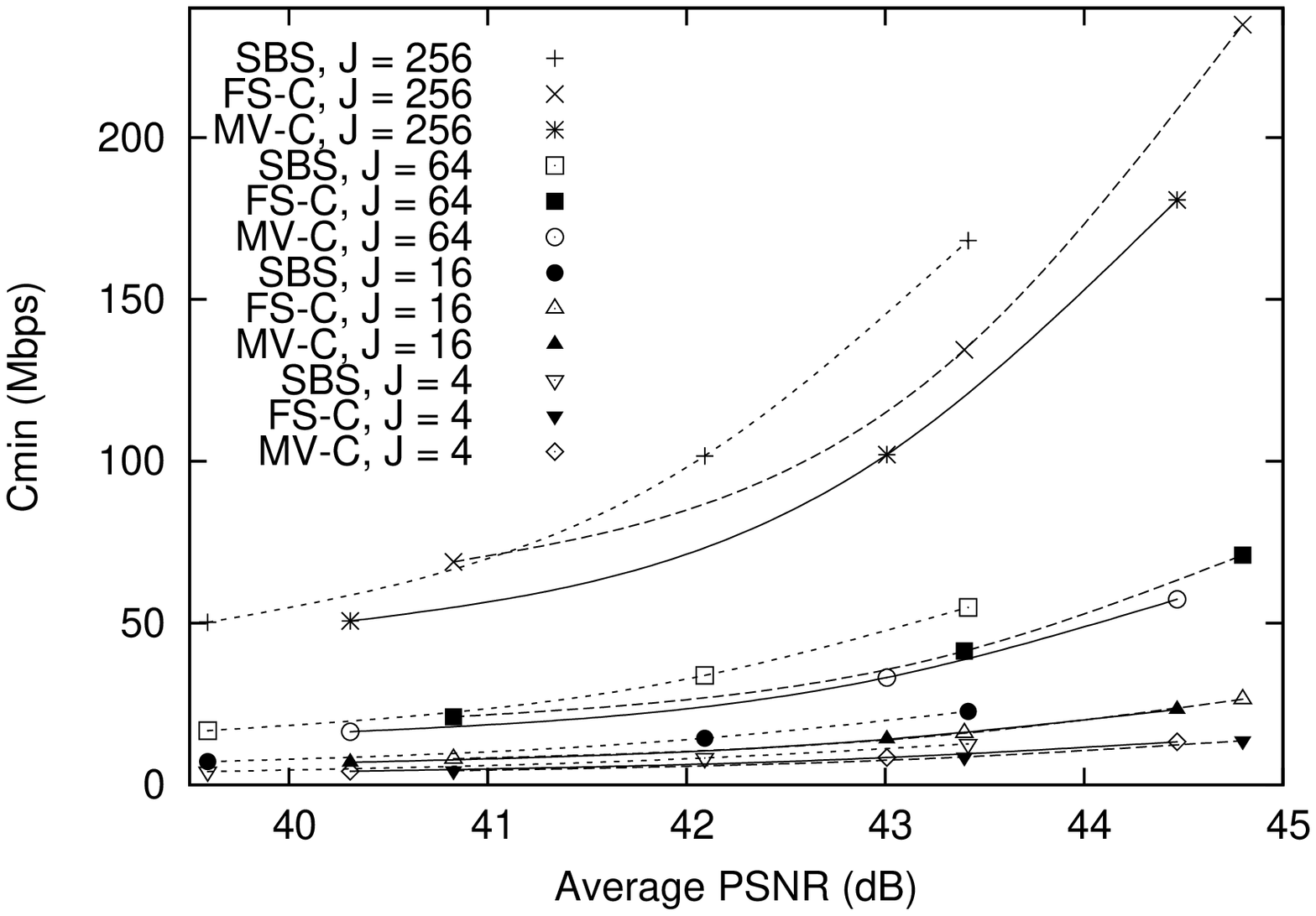} &
      \includegraphics[height=.34\textheight,angle=270]{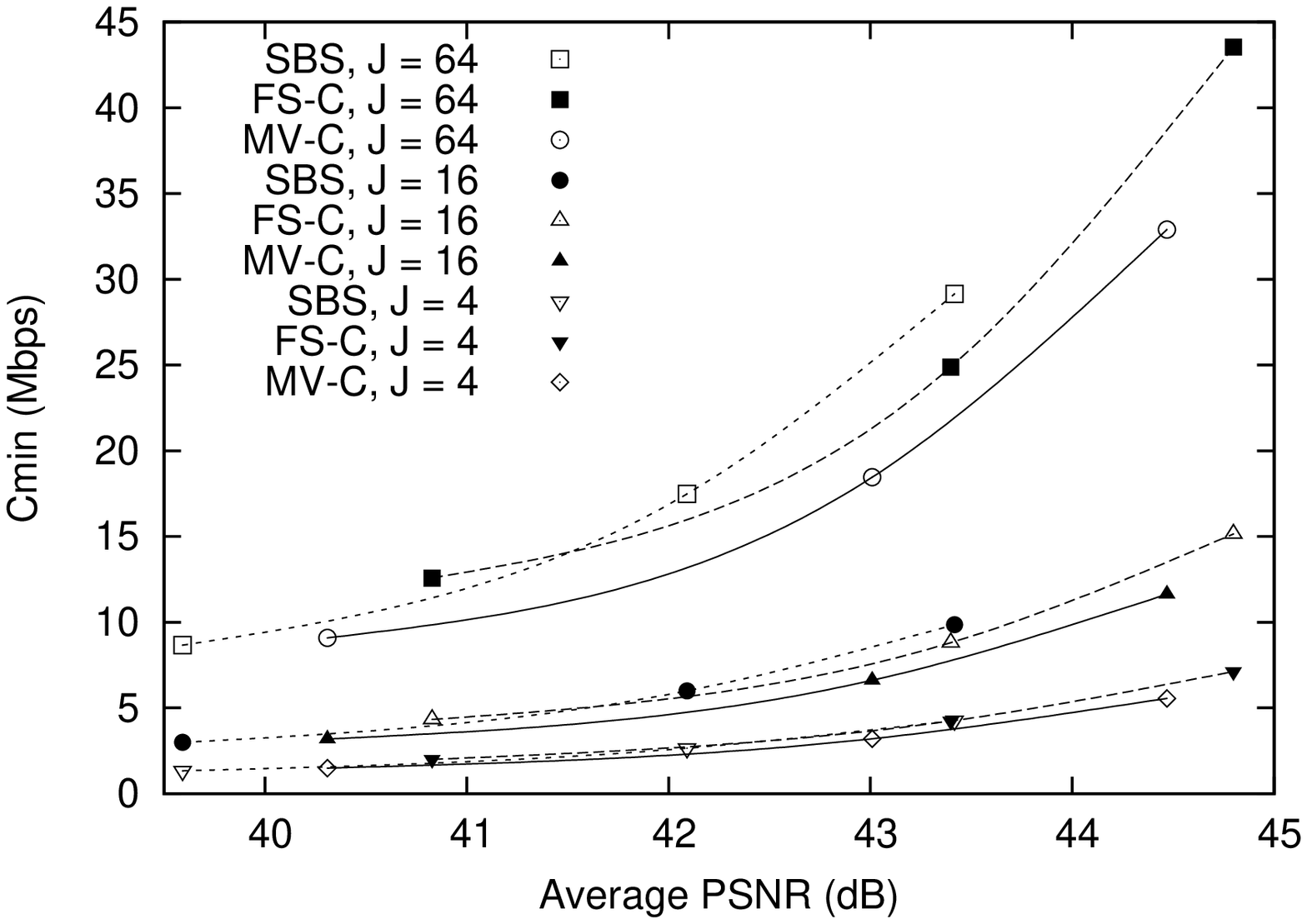} \\
           {\footnotesize (c) \textit{Alice in Wonderland}, frame-by-frame} &
           {\footnotesize (d) \textit{Alice in Wonderland}, GoP smoothing} \\
      \includegraphics[height=.34\textheight,angle=270]{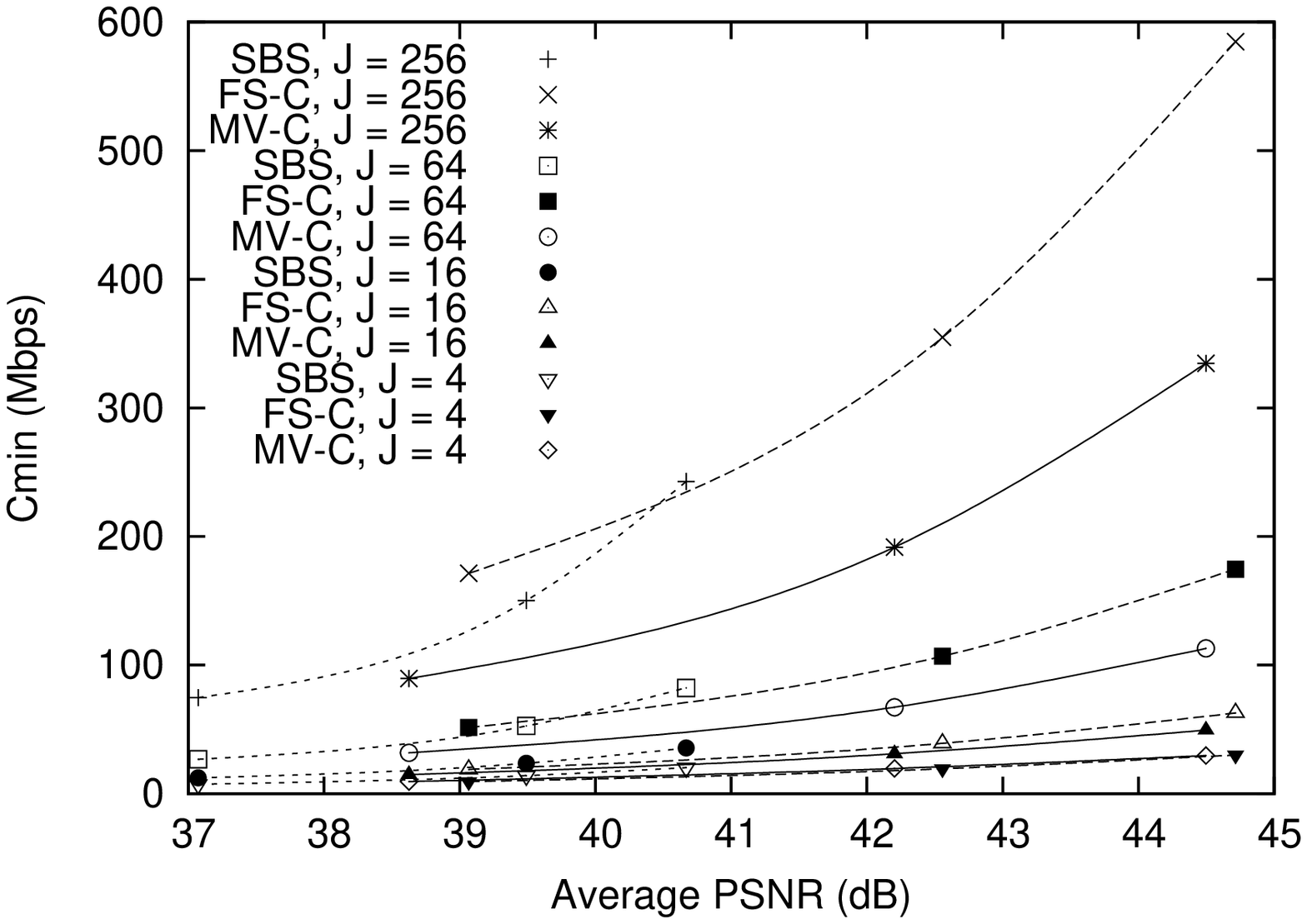} &
      \includegraphics[height=.34\textheight,angle=270]{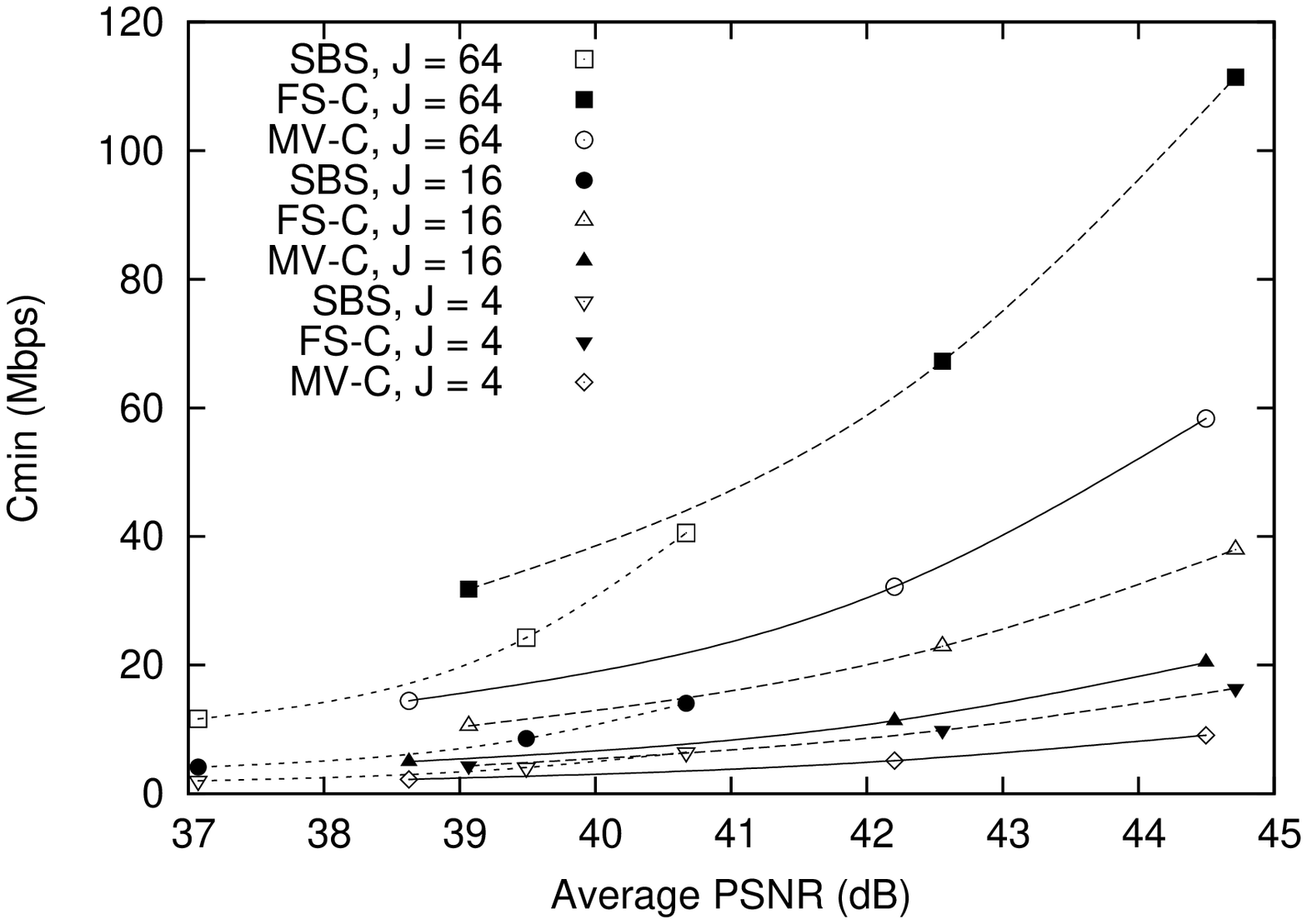} \\
           {\footnotesize (e) \textit{IMAX Space Station}, frame-by-frame} &
           {\footnotesize (f) \textit{IMAX Space Station}, GoP smoothing} \\
\end{tabular}
    \caption{Required minimum link transmission bit rate $C_{\min}$
    to transmit $J$ streams with an information loss probability
   $P_{\rm loss}^{\rm info} \leq \epsilon = 10^{-5}$.
    GoP structure B7 with seven B frames between I and P frames.}
    \label{Cminall}
\end{figure*}

\begin{figure*}[!htb]
 \begin{tabular}{cc}
  \includegraphics[height=.34\textheight,angle=270]{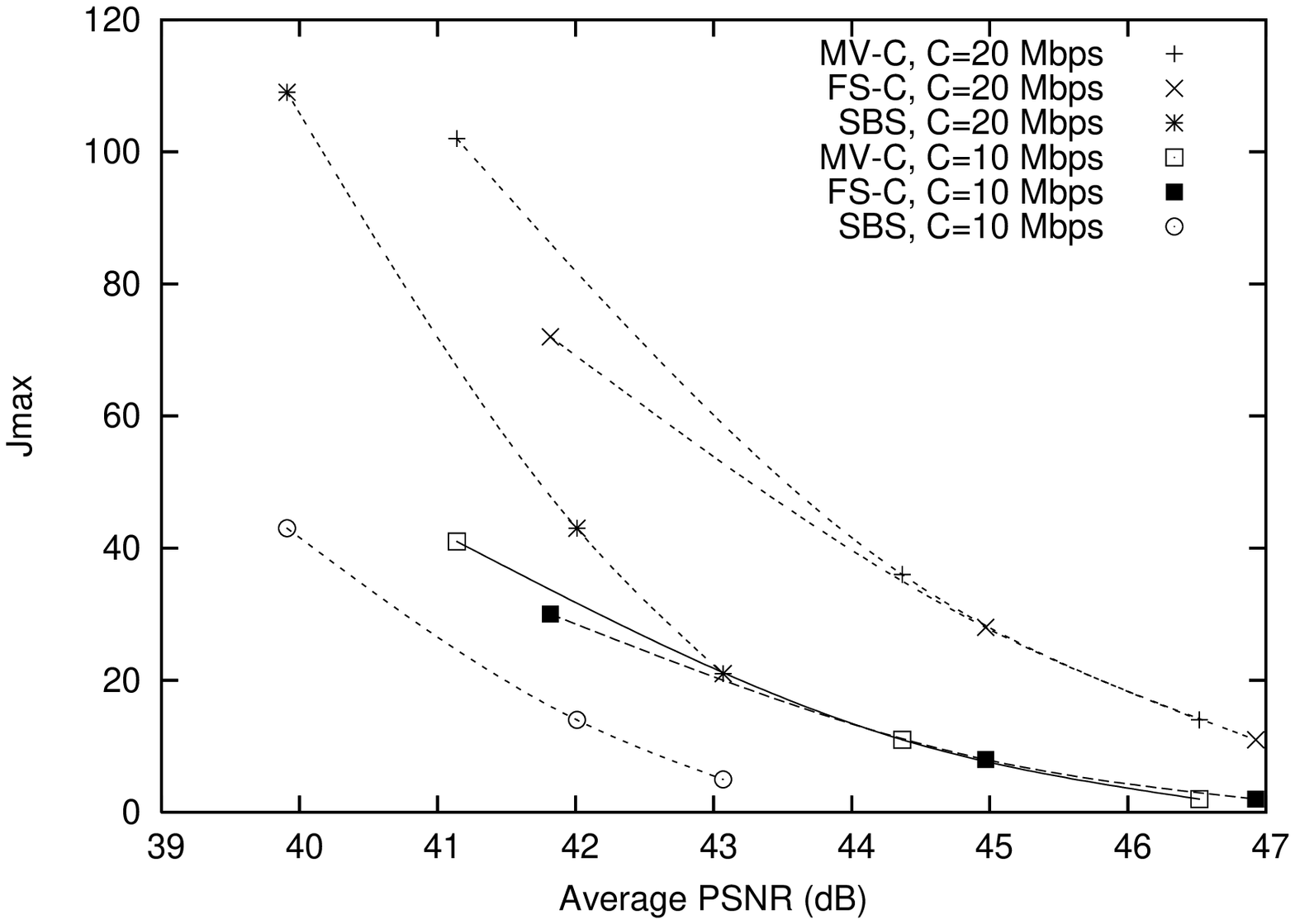} &
  \includegraphics[height=.34\textheight,angle=270]{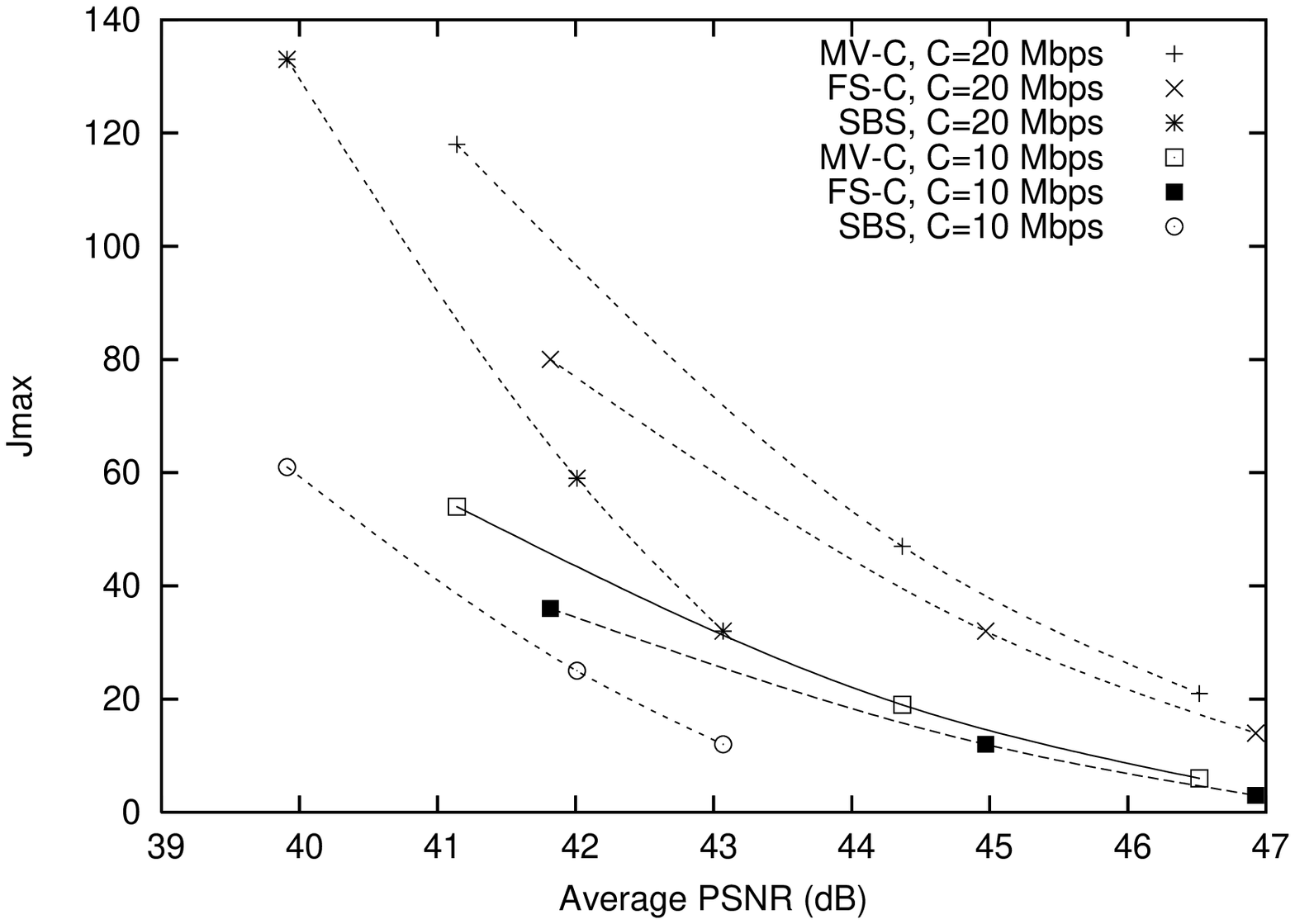} \\
  {\footnotesize (a) \textit{Monsters vs Aliens}, frame-by-frame} &
  {\footnotesize (b) \textit{Monsters vs Aliens}, GoP smoothing} \\
      \includegraphics[height=.34\textheight,angle=270]{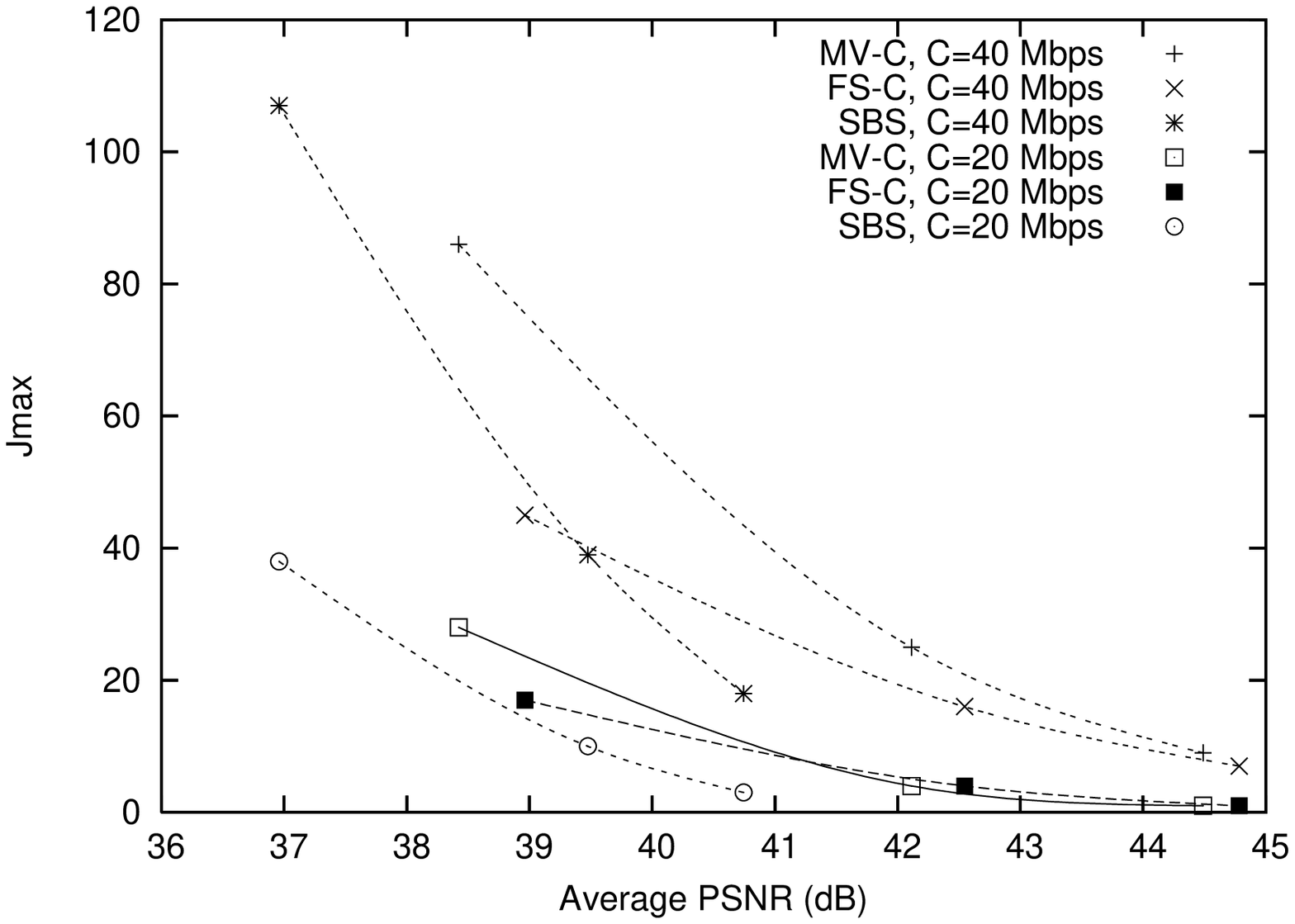} &
      \includegraphics[height=.34\textheight,angle=270]{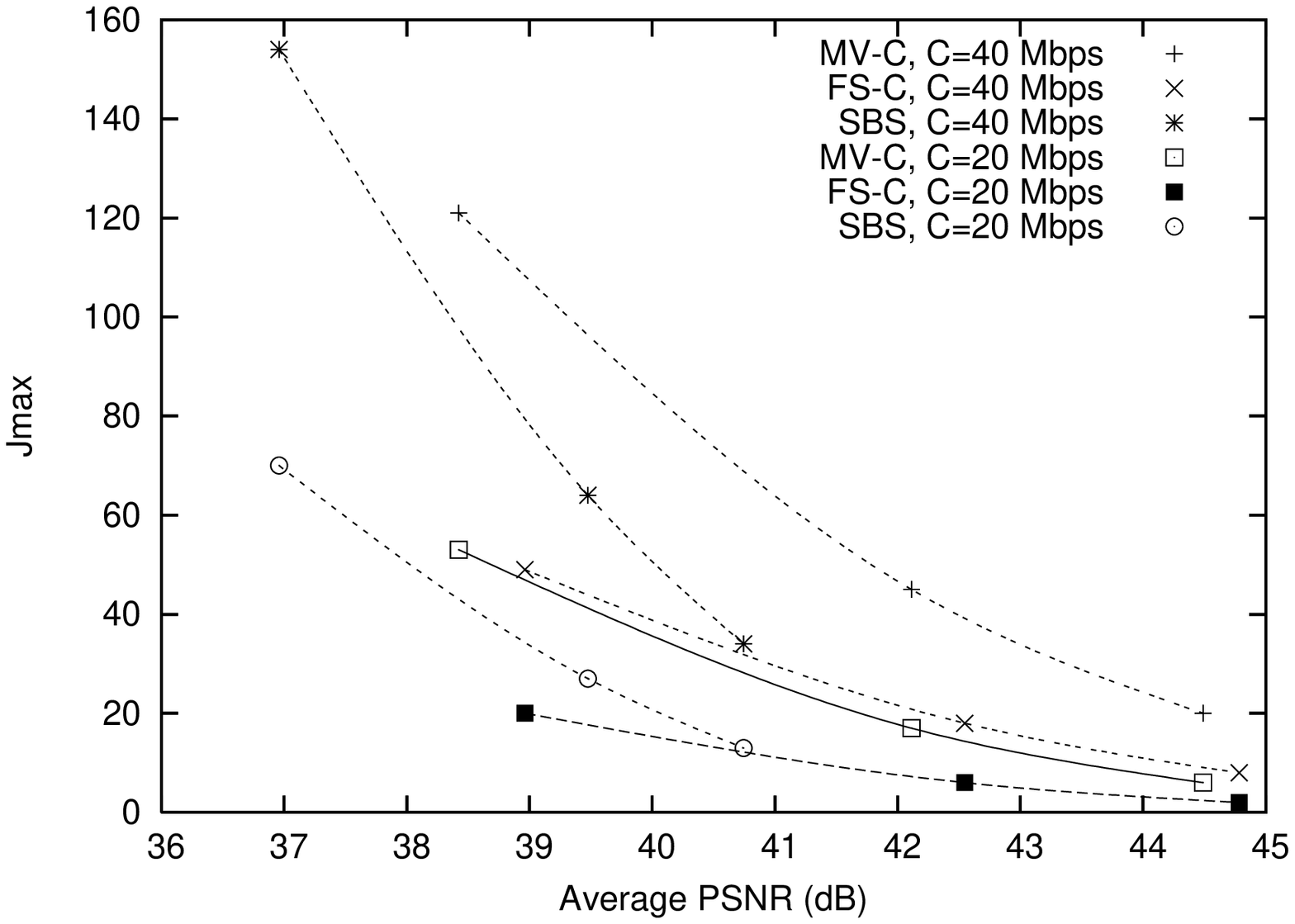} \\
           {\footnotesize (c) \textit{IMAX Space Station}, frame-by-frame} &
           {\footnotesize (d) \textit{IMAX Space Station}, GoP smoothing} \\
\end{tabular}
    \caption{Maximum number of supported streams with 
    an information loss probability
   $P_{\rm loss}^{\rm info} \leq \epsilon = 10^{-5}$ for given link
    transmission bit rate $C$.
   GoP structure B1 with one B frame between I and P frames.}
    \label{JmaxallB1}
\end{figure*}
\begin{figure*}[!htb]
 \begin{tabular}{cc}
  \includegraphics[height=.34\textheight,angle=270]{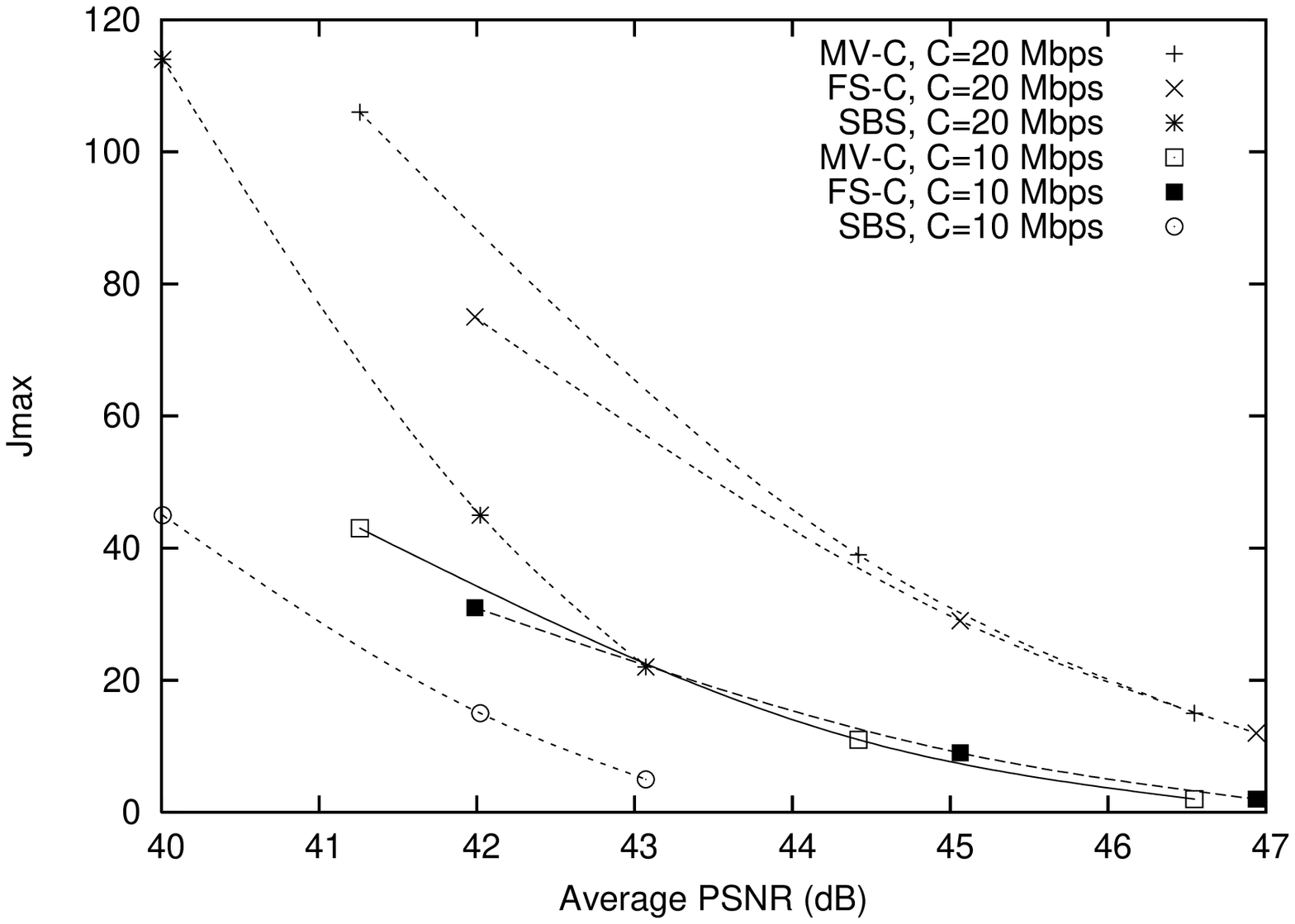} &
  \includegraphics[height=.34\textheight,angle=270]{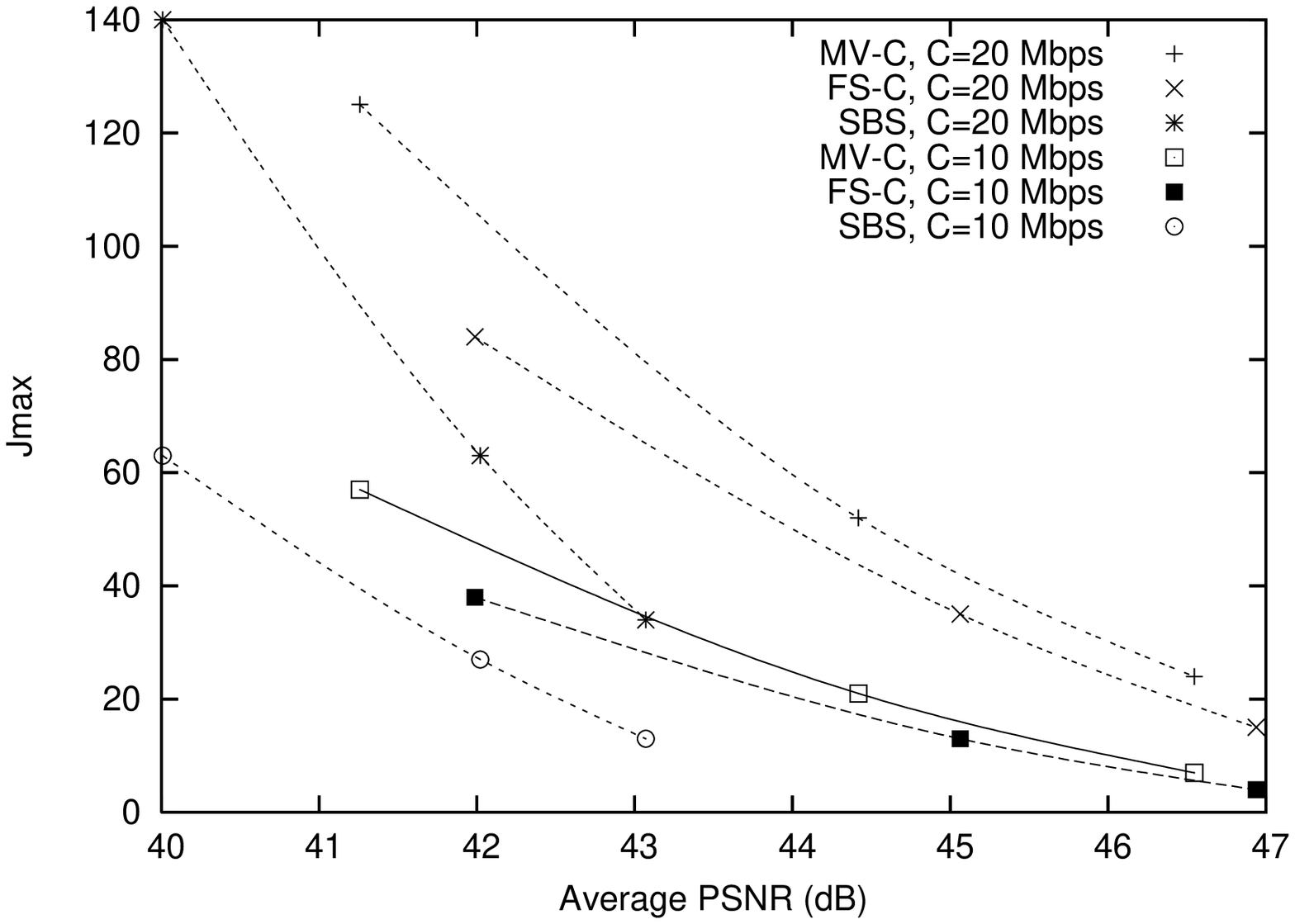} \\
  {\footnotesize (a) \textit{Monsters vs Aliens}, frame-by-frame} &
  {\footnotesize (b) \textit{Monsters vs Aliens}, GoP smoothing} \\
      \includegraphics[height=.34\textheight,angle=270]{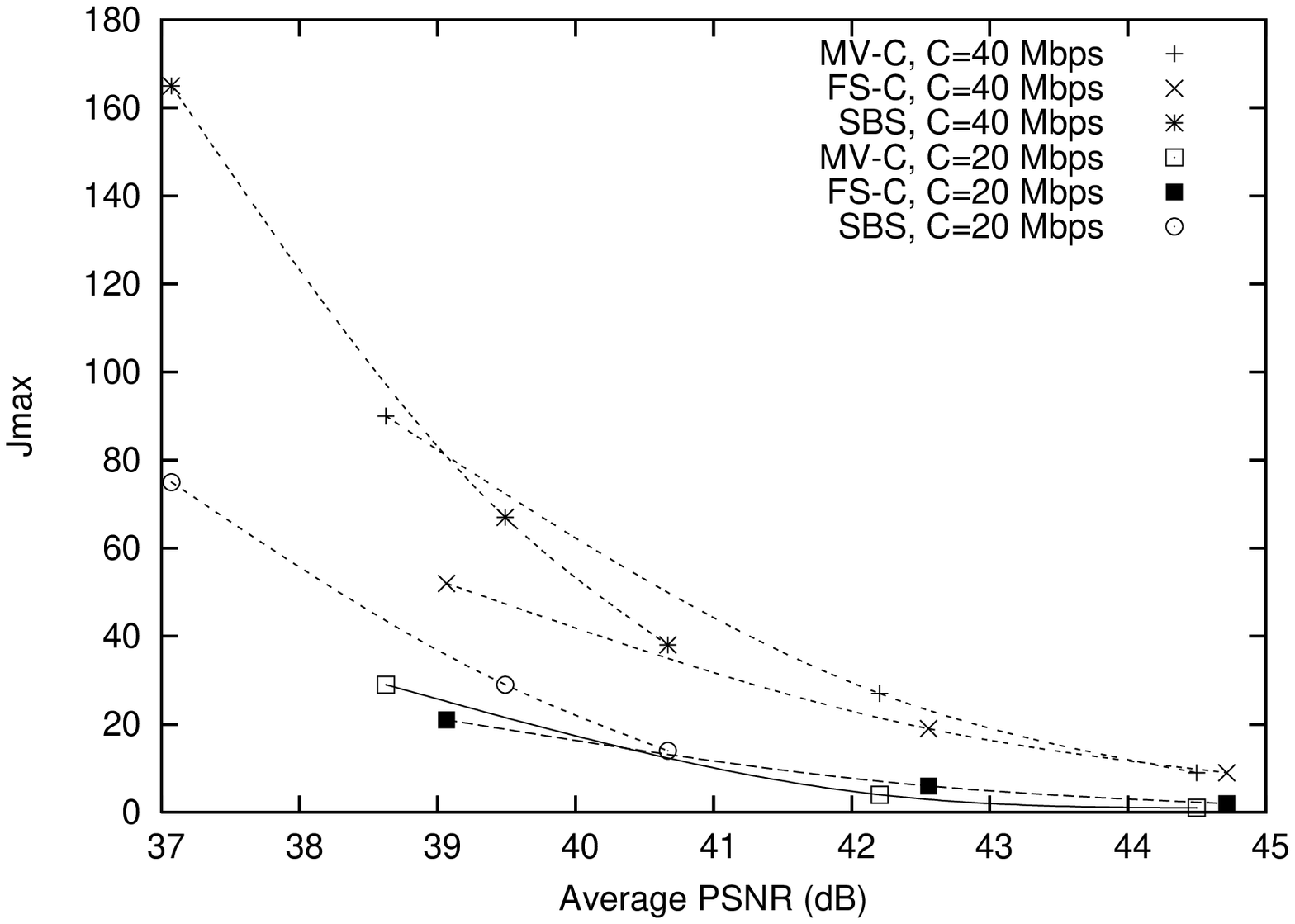} &
      \includegraphics[height=.34\textheight,angle=270]{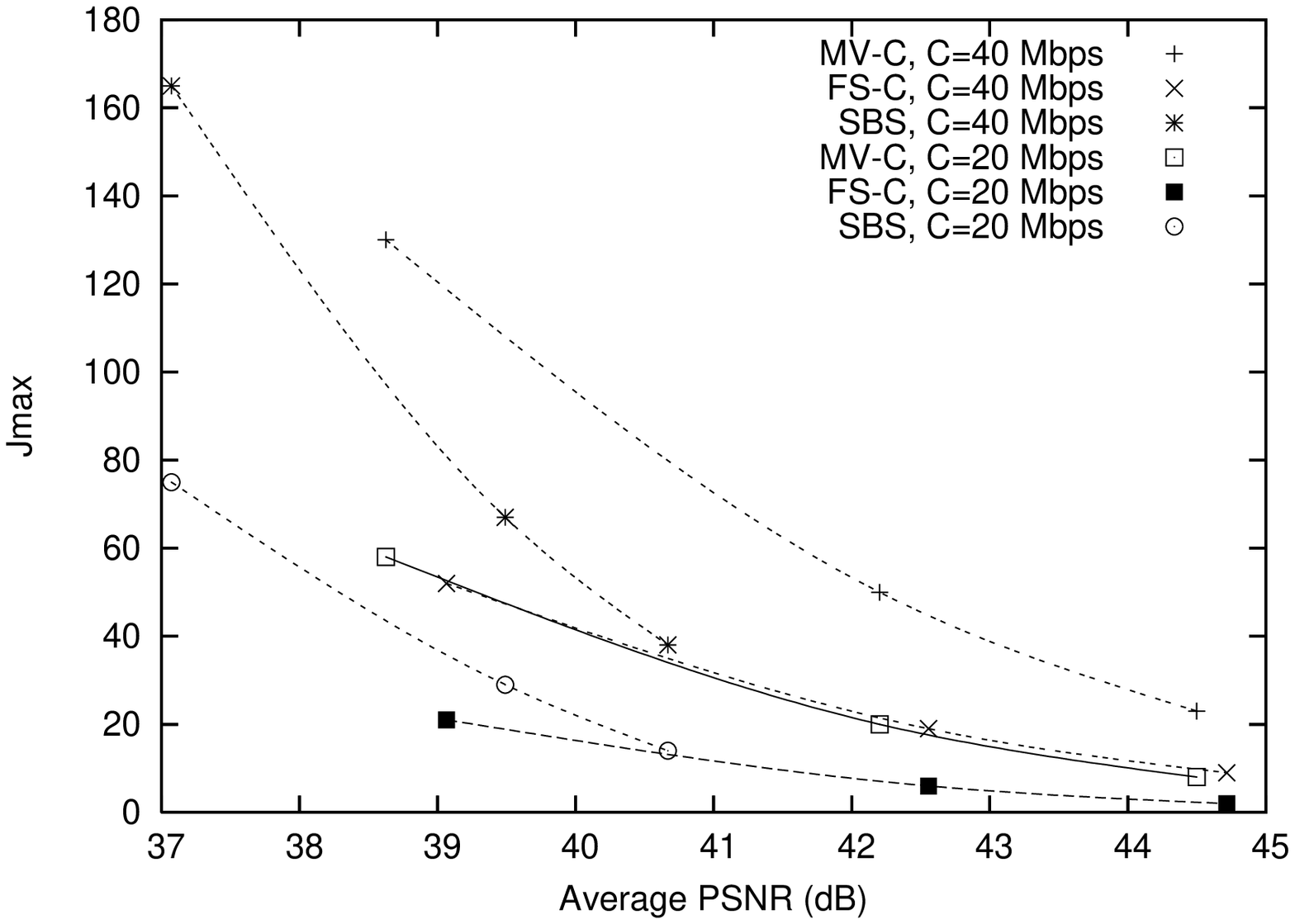} \\
           {\footnotesize (c) \textit{IMAX Space Station}, frame-by-frame} &
           {\footnotesize (d) \textit{IMAX Space Station}, GoP smoothing} \\
\end{tabular}
    \caption{Maximum number of supported streams with 
    an information loss probability
   $P_{\rm loss}^{\rm info} \leq \epsilon = 10^{-5}$ for given link
    transmission bit rate $C$.
   GoP structure B7 with seven B frames between I and P frames.}
    \label{JmaxallB7}
\end{figure*}
\subsubsection{Simulation Setup}
We consider a single ``bufferless'' statistical
multiplexer~\cite{RaJFA05,RoMV96,paper46}
which reveals the fundamental statistical multiplexing behaviors
without introducing arbitrary parameters, such as buffer sizes,
cross traffic, or multi-hop routing paths.
Specifically, we consider a link of transmission bitrate
$C$ [bit/s], preceded by a buffer of size $C/f$ [bit], i.e.,
the buffer holds as many bits as can be transmitted in one
frame period of duration $1/f$.
We let $J$ denote the number of 3D video streams
fed into the buffer.
Each of the $J$ streams in a given simulation is derived from the same encoded
3D video sequence; whereby, each stream has its own starting frame
that is
drawn independently, uniformly, and randomly from the set $[1, 2, \ldots, \ M]$.
Starting from the selected starting frame, each of the $J$ videos
places one encoded frame
of the SBS representation (multiview frame of the MV-C or FS-C representation)
into the multiplexer buffer in each
frame period.
If the number of bits placed in the buffer in a frame period exceeds
$C/f$, then there is loss.
We count the number of lost bits to evaluate the information loss probability
$P_{\rm loss}^{\rm info}$~\cite{RoMV96}
as the proportion of the number of lost bits to the number
of bits placed in the multiplexer buffer.
We conduct many independent replications of the stream multiplexing, each
replication simulates the transmission of $M$ frames
(with ``wrap-around'' to the first frame when the end of the video is reached)
for each stream, and each replication has
a new independent set of random starting frames for the $J$
streams.

\subsubsection{Evaluation Results}
We conducted two types of evaluations. First, we determined
the maximum number of streams $J_{\max}$ that can be transmitted over the
link with prescribed transmission bit rate $C = 10,\ 20$, and 40 Mb/s
such that
$P_{\rm loss}^{\rm info}$ is less than a prescribed small
$\epsilon = 10^{-5}$.
We terminated a simulation when
the confidence interval for $P_{\rm loss}^{\rm info}$
was less than 10~\% of the corresponding sample mean.

Second, we estimated the
minimum link capacity $C_{\min}$ that accommodates a prescribed
number of streams $J$ while keeping
$P_{\rm loss}^{\rm info} \leq \epsilon = 10^{-5}$.
For each $C_{\min}$ estimate, we performed 500 runs,
each consisting of 1000 independent video streaming simulations.
We discuss in detail the representative results
from this evaluation of $C_{\min}$ for a given number of streams $J$.
The results for the evaluation of $J_{\max}$ given a fixed
link capacity $C$ indicate the same tendencies.

We observe from Figs.~\ref{Cminall}(a), (c), and (e) that for
small numbers of multiplexed streams, namely $J = 4$ and 16 streams
for \textit{Monsters vs Aliens} and \textit{Alice in Wonderland}, as well as
$J = 4$ streams for
\textit{IMAX Space Station}, the MV and FS representations require
essentially the same transmission bitrate.
Even though the MV representation and encoding has higher
RD efficiency, i.e., lower average bit rate for a given
average PSNR video quality, the higher MV traffic variability makes
statistical multiplexing more challenging, requiring the same
transmission bit rate as the less RD efficient FS representation
(which has lower traffic variability).
We further observe from Figs.~\ref{Cminall}(a), (c) and (e) that
increasing the statistical multiplexing effect by multiplexing
more streams, reduces the effect of the traffic variability, and,
as a result, reduces the required transmission bit rate $C_{\min}$
for MV-C relative to FS-C.

We observe from Figs.~\ref{Cminall}(b), (d), and (f) that
GoP smoothing effectively reduces the MV traffic variability such that
already for small numbers of multiplexed streams, i.e., a weak
statistical multiplexing effect, the required transmission bitrate for MV is
less than that for FS.

\section{Conclusion and Future Work}
\label{con} We have compared the traffic characteristics and
fundamental statistical multiplexing behaviors of state-of-the-art
multiview (MV) 3D video representation and encoding with the frame
sequential (FS) and side-by-side (SBS) representations encoded with
state-of-the-art single-view video encoding. We found that
the SBS representation, which permits transmission of two-view video
with the existing single-view infrastructure, incurs significant
PSNR quality degradations compared to the MV and FS representations
due to the sub-sampling and interpolation involved with the SBS
representation. We found that when transmitting small numbers of
streams without traffic smoothing, the higher traffic variability of
the MV encoding leads to the same transmission bitrate requirements
as the less RD efficient FS representation with single-view coding.
We further found that to reap the benefit of the more RD efficient
MV representation and coding for network transport, traffic
smoothing or the multiplexing of many streams in large transmission
systems is required.

There are many important directions for future research on the
traffic characterization
and efficient network transport of encoded 3D video, and generally
multiview video. One direction is to develop and evaluate
smoothing and scheduling mechanisms that consider a wider set
of network and receiver constraints, such as limited
receiver buffer or varying wireless link bitrates, or
collaborate across several ongoing streams~\cite{BaLi02,ReRo98}.
Another avenue is to exploit network and client resources, such
as caches or cooperating peer clients for efficient delivery of
multiview video services.
Broadly speaking, these effective transmission strategies
are especially critical when relatively few video streams
are multiplexed as, for instance, 
in access networks, e.g.,~\cite{AuSR11,SKM0110,ZhMo09},
and metro networks, e.g.,~\cite{BiBC13,MaRe04,ScMRW03,YuCL10}.
Moreover, the challenges are especially pronounced in
networking scenarios in support of applications with
tight real-time constraints, such as 
gaming~\cite{BrF10,FiGR02,ScER02} and real-time
conferencing and tele-immersion~\cite{KuB13,PaDR12,VaZK10}.

\bibliographystyle{IEEEtran}



\end{document}